\documentclass[epj,final]{svjour}
\usepackage{graphicx}
\begin{document}
\title{Pionic deuterium}
\author{Th. Strauch\inst{1},
        F.\,D.\,Amaro\inst{2}, 
        D.\,F.\,Anagnostopoulos\inst{3},
        P.\,B\"uhler\inst{4}, 
        D.\,S.\,Covita\inst{2,7}\thanks{\emph{present address:} 
        I3N, Dept. of Physics, Aveiro University, P-3810 Aveiro, Portugal}, 
        H.\,Gorke\inst{5}, 
        D.\,Gotta\inst{1}\thanks{\emph{corresponding author:} d.gotta@fz-juelich.de}, 
        A.\,Gruber\inst{4}, 
        A.\,Hirtl\inst{4}\thanks{\emph{present address:} 
        Universit\"atsklinik f\"ur Nuklearmedizin, Medizinische Universit\"at Wien, 1090 Vienna, Austria}, 
        P.\,Indelicato\inst{6},
        E.-O.\,Le Bigot\inst{6},
        M.\,Nekipelov\inst{1},
        J.\,M.\,F.\,dos Santos\inst{2},
        Ph.\,Schmid\inst{4},
        S.\,Schlesser\inst{6},
        L.\,M. Simons\inst{7},
        M.\,Trassinelli\inst{6}\thanks{\emph{present address:} 
        Inst. des NanoSciences de Paris, CNRS UMR7588 and UMPC-Paris 6, F-75015 Paris, France},
        J.\,F.\,C.\,A.\,Veloso\inst{8},
   \and J.~Zmeskal\inst{4}
}                     
%
%
\institute{     Institut f\"ur Kernphysik, Forschungszentrum J\"ulich, D-52425 J\"ulich, Germany
           \and Dept. of Physics, Coimbra University, P-3000 Coimbra, Portugal
           \and Dept. of Materials Science and Engineering, University of Ioannina, GR-45110 Ioannina, Greece 
           \and Stefan Meyer Institut for Subatomic Physics, Austrian Academy of Sciences, A-1090 Vienna, Austria
           \and Zentralinstitut f\"{u}r Elektronik, Forschungszentrum J\"{u}lich GmbH, D-52425 J\"{u}lich, Germany
           \and Laboratoire Kastler Brossel, UPMC-Paris 6, ENS, CNRS; Case 74, 4 place Jussieu, F-75005 Paris, France 
           \and Laboratory for Particle Physics, Paul Scherrer Institut, CH-5232 Villigen, Switzerland
           \and I3N, Dept. of Physics, Aveiro University, P-3810 Aveiro, Portugal}
\date{Received: date / Revised version: date}
%
\abstract{
The strong-interaction shift $\epsilon^{\pi\mathrm{D}}_{1s}$ and broadening $\Gamma^{\pi\mathrm{D}}_{1s}$ in pionic deuterium 
have been determined in a high statistics study of the $\pi$D$(3p-1s)$ X-ray transition using a high-resolution 
crystal spectrometer. The pionic deuterium shift will provide constraints for the pion-nucleon isospin scattering 
lengths extracted from measurements of shift and broadening in pionic hydrogen. The hadronic broadening is related 
to pion absorption and production at threshold. The results are 
\vspace{0.75mm}\linebreak 
$\epsilon^{\pi\mathrm{D}}_{1s}=(-2356\pm 31)$\,meV (repulsive) and $\Gamma^{\pi\mathrm{D}}_{1s}=(1171{+\,23\atop-\,49})$\,meV 
yielding for the complex $\pi$D scattering 
\vspace{-1.0mm}\linebreak 
length $a_{\pi\mathrm{D}}=[-(24.99\pm 0.33) +\,i\,(6.22{+\,0.12\atop -0.26})]\times 10^{-3}\,m^{-1}_{\pi}$.  
From the imaginary part, the threshold 
\vspace{-1.5mm}\linebreak
parameter for pion production is obtained to be $\alpha=(251{+\,5\atop -11})\,\mu$b. This allows, in addition, and by using 
\vspace{-1.5mm}\linebreak
results from pion absorption in $^3$He at threshold, the determination of the effective couplings $g_0$ and $g_1$ for 
s-wave pion absorption on isoscalar and isovector $NN$ pairs.
} 
\authorrunning{Th.\,Strauch {\it et al.}}
\titlerunning{Pionic Deuterium}
\maketitle
\section{Introduction}\label{sec:intro}

Hadronic atoms reveal the influence of the strong force by a shift $\epsilon$ and broadening 
$\Gamma$ of the low-lying atomic levels with respect to the pure electromagnetic interaction. 
As atomic binding energies are negligibly small compared to the hadronic scale, a measurement of shift and 
width is equivalent to a scattering experiment at zero energy (threshold). Hence, such atomic 
data contain information on hadron-nucleus scattering lengths\,\cite{Des54,Got04}. 

In pionic hydrogen, the atomic ground state level shift $\epsilon^{\pi\mathrm{H}}_{1s}$ and broadening $\Gamma^{\pi\mathrm{H}}_{1s}$ 
are connected to the two isospin-separated pion-nucleon ($\pi N$) scattering lengths. In the case of nuclei $A(N,Z)$ 
with $A\geq 2$, $\epsilon^{\pi\mathrm{A}}_{1s}$ and $\Gamma^{\pi\mathrm{A}}_{1s}$ are measures of the complex pion-nucleus scattering 
length $a_{\pi\mathrm{A}}$. The real part of $a_{\pi\mathrm{A}}$ is attributed in leading order to scattering, whereas the imaginary 
part is due to pion absorption inducing itself a contribution to the shift\,\cite{Eri88-6,Bru55}. 

It has been shown, that strong-interaction effects in low $Z$ pionic atoms can be related unambiguously to pion-nucleon ($\pi N$) 
scattering lengths as defined in modern field theoretical approaches of QCD\,\cite{Gas08}. Such threshold quantities are of great 
importance because they belong to the first scattering parameters accessible by lattice calculations\,\cite{Lue86}. 
For the $\pi N$ case, however, results are not yet available\,\cite{Tor10,Bea11}.

Pion-nucleus dynamics is understood to be built up from elementary $\pi N\rightarrow \pi N$ and $\pi NN\leftrightarrow NN$ processes 
taking into account nuclear structure, multiple scattering, and absorptive phenomena. Vice versa, precise pion-nucleus data could 
set constraints on elementary amplitudes. Generally, systematic uncertainties involved in the extraction of 
$a_{\pi\mathrm{A}}$ from atom data, if available, are smaller than in the extrapolation of cross-section data to threshold. 
In this case, normalisation problems cease to exist and corrections owing to Coulomb bound states are better under 
control than for scattering states (see sect.\,\ref{strong_interaction}).

Assuming isospin conservation and charge symmetry, $\pi N\rightarrow \pi N$ scattering at threshold is described 
completely by two real numbers, {\it e.g.}, the isoscalar and isovector $\pi N$ scattering lengths $a^+$ and $a^-$. They 
are defined in terms of two $\pi N\rightarrow \pi N$ reactions or isospin $I=1/2$ and $I=3/2$ by 
\begin{eqnarray}
a^{+}&=&\frac{1}{2}\,\,(a_{\pi^{-}p\to\pi^{-}p}+a_{\pi^{+}p\to\pi^{+}p})=\frac{1}{3}\,(a_{1/2}+2a_{3/2})\,,\nonumber\\
a^{-}&=&\frac{1}{2}\,\,(a_{\pi^{-}p\to\pi^{-}p}-a_{\pi^{+}p\to\pi^{+}p})=\frac{1}{3}\,(a_{1/2}-a_{3/2})\label{eq:apm}.
\end{eqnarray} 
Furthermore, among others the relations $a_{\pi^{-}n\to\pi^{-}n}=a_{\pi^{+}p\to\pi^{+}p}$ and 
$a_{\pi^{-}p\to\pi^{\circ}n}= -\,\sqrt{2}\,a^{-}$ hold.

The scattering lengths $a^+$ and $a^-$ play a key role in the modern low-energy approach of QCD based on 
effective field theories (EFT). A perturbative method---chiral perturbation theory ($\chi$PT)---has been 
developed, which is intimately related to the smallness of the pion's mass $m_{\pi}$ due to its underlying 
nature as a Goldstone boson\,\cite{Wei66,Tom66,Gas82,ThoWei}. The interaction is calculated order by order as 
an expansion in the---compared to the hadronic scale of 1\,GeV---"small" quantities momentum, pion mass, fine 
structure constant thus including the 
(isospin--breaking) electromagnetic interaction, and the light quark mass difference $m_{d}-m_{u}$ (strong 
isospin breaking)\,\cite{Wei79,Gas84,Gas85,Eck95,Ber95,Sch02,Ber08}. 
In such a theory short--range contributions are summarized in so called low-energy constants (LECs) which must be 
determined from $\pi N$ experiments (for details see\,\cite{Gas08}). 

From the scattering lengths $a^+$ and $a^-$, important quantities of the $\pi N$ interaction are derived\,\cite{Bec01}. 
The $\pi N$ $\sigma$\,term\,\cite{ThoWei,Gas91,Sai02} is a measure of explicit chiral symmetry breaking, and the 
$\pi N$ coupling constant, obtained from the Gold\-berger-Miyazawa-Oehme sum rule\,\cite{Gol55,Eri02,Aba07}, is related 
to the strength of threshold photo production of charged pions\,\cite{Ber96,Kov97,Han97} and the induced pseudoscalar 
coupling $g_p$ as determined from muon capture in hydrogen\,\cite{Mea01,Gor04,And07}.

Access to $a^+$ and $a^-$ is available from the $1s$ level shift and broadening in pionic hydrogen ($\pi$H):
\begin{eqnarray}
\epsilon^{\pi\mathrm{H}}_{1s}&\propto & a_{\pi^{-}p\to\pi^{-}p}\hspace{19mm}=\,a^{+}+a^{-}~+~...\\
\Gamma^{\pi\mathrm{H}}_{1s}  &\propto & \left(1+\frac{1}{P}\right)(a_{\pi^{-}p\to\pi^{0}n})^2\propto\,(a^{-})^2\,~~~~+~...\,.\label{eq:eGpiH}
\end{eqnarray} 
Ellipses stand for corrections owing to the fact that the pionic atom constitutes a Coulomb state as well as for electromagnetic and strong 
isospin and non-isospin breaking corrections, which are essential to extract $a^+$ and $a^-$ in a well defined way from the data. These corrections  
may be calculated, {\it e.g.}, within the framework of $\chi$PT and are of the order of a few per cent\,\cite{Gas08,Sig96b,Lyu00,Gas02,Zem03}. 

In the $\pi$H case, the level width is as well due to scatter\-ing---the charge exchange reaction $\pi^{-}p\to\pi^{\circ}n$---and, 
therefore, is proportional to the square of $a^{-}$\,\cite{Ras82}. The Panofsky ratio 
$P=\Gamma(\pi^{-}p\to\pi^{\circ}n)/\Gamma(\pi^{-}p\to\gamma n)=1.546\pm 0.009$\,\cite{Spu77} accounts for 
the relative strength of radiative capture, the second channel contributing to the total broadening. For a recent 
analysis of $\pi$H atomic data see ref.\,\cite{Oad07}.

It is known already from current algebra, that $a^{+}$ vanishes in the chiral limit \,\cite{Wei66,Tom66} 
acquiring a finite value only because of chiral symmetry breaking\,\cite{ThoWei,Eri88-9}. Hence, 
the s-wave $\pi N$ interaction is dominated by the isovector contribution with the consequence that it is 
difficult to extract $a^{+}$ from pion-nucleon and nucleus data with $I\neq 0$. 

Pionic deuterium ($\pi$D) is of particular interest being the first natural choice to test the interaction of 
pions with isoscalar nuclear matter. Here, the leading order (LO) one-body contribution to the real part of the 
pion-deuteron scattering length $a_{\pi\mathrm{D}}$ depends only on the isoscalar quantity $a^+$. Considering the 
deuteron as a free proton and neutron plus corrections, one may write  
\begin{eqnarray}
\epsilon^{\pi\mathrm{D}}_{1s}&\propto&\mathrm{Re}\,a_{\pi\mathrm{D}}\nonumber\\
                     &=&a_{\pi^{-}p\to\pi^{-}p}+a_{\pi^{-}n\to\pi^{-}n}~+~...\nonumber\\
                     &=&\hspace{15mm}2a^{+}\hspace{15mm}~+~...\label{eq:epiD}\,.
\end{eqnarray} 
Besides the corrections mentioned for the case of $\pi$H, ellipses include here terms arising from multiple 
scattering, depending both on $a^+$ and $a^-$, and absorptive contributions. Noteworthy, that the second order 
correction, dominated by $(a^{-})^2$, exceeds even the magnitude of the leading order term because 
of $a^+\ll a^-$\,\cite{Wei66,Tom66}. Nuclear structure is taken into account by folding with the deuteron wave 
function\,\cite{Eri88-4,ThLa80,Del03}. The exact relation between $\epsilon^{\pi\mathrm{D}}_{1s}$ and 
$\mathrm{Re}\,a_{\pi\mathrm{D}}$ is discussed in detail in sect.\,\ref{sec:strong_interaction_shift}. 

Hence, the isopin scattering lengths are accessible from the shift measurements in $\pi$H and in $\pi$D already 
without ultimately precise data for the hadronic broadening in hydrogen being available\,\cite{Eri02}. An effective  
constraint on $a^+$ and $a^-$, however, is obtained when combining the triple $\epsilon^{\pi\mathrm{H}}_{1s}$, 
$\Gamma^{\pi\mathrm{H}}_{1s}$, and $\epsilon^{\pi\mathrm{D}}_{1s}$. A complete set of data has been achieved for the first time 
during the last decade\,\cite{Sig96a,Cha9597,Hau98,Sch01}. 
 
Such a constraint is highly desirable, because different methods exist to obtain the corrections needed for the 
extraction of the pure hadronic quantities $a^+$ and $a^-$ from $\pi$H and $\pi$D measurements. Recognising, among 
others, that  corrections obtained using potential models were incomplete, significant effort went into calculations 
to establish an unambiguous relation between atomic data and $a^+$ and $a^-$, in particular within the framework of 
$\chi$PT. It has been proven to be extendable in a defined way to 3-body interactions\,\cite{Wei92} and allows to 
treat electromagnetic and strong isopin-breaking terms on the same footing 
\cite{Gas08,Eri02,Bar97,Bea98,Tar00,Del01,Kai02,Bur03,Bea03,Doe04,Irg04,Mei05,Mei06,Hof09a}.
For a discussion and comparison of potential and EFT approaches see ref.\,\cite{Gas08}.

As mentioned above, corrections for the $\pi N$ scattering lengths as obtained from $\pi$H amount to a few per cent 
of the leading contribution  in complete contrast to the real part of the $\pi$D scattering length. Here, because of 
the smallness of $a^+$, the isospin-breaking contribution was found to be as large as 40\% in NLO $\chi$PT\,\cite{Mei06}.

Finally, it turned out that for the extraction of the scattering lengths $a^{+}$ from the atomic data within the framework 
of $\chi$PT, the accuracy achievable is determined by one particular combination of LECs. It appears both in the 
correction term for $\epsilon^{\pi\mathrm{H}}_{1s}$ and $\epsilon^{\pi\mathrm{D}}_{1s}$, and thus constitutes a limit from the theoretical 
side\,\cite{Bar11}. In the case of $\Gamma^{\pi\mathrm{H}}_{1s}$ the LECs involved are known better\,\cite{Gas08,Hof09b}. Therefore, 
here the uncertainty of $a^{-}$ is dominated by the experimental accuracy.

Being different from the $\pi$H case, the hadronic broadening in $\pi$D and, consequently, the imaginary part of the 
$a_{\pi\mathrm{D}}$ is not related to $\pi N$ scattering lengths $a^+$ and $a^-$ but to the strength of s-wave pion absorption. 
Open channels in negative pion absorption in deuterium at threshold are 
\begin{eqnarray}
\pi^-d&\rightarrow &nn\label{eq:nn}\\
      &\rightarrow &nn\gamma\label{eq:nng} \\
      &\rightarrow &nne^+e^-\label{eq:nnee}\\
      &\rightarrow &nn\pi^0\label{eq:nnpi0}\,. 
\end{eqnarray}
True absorption (\ref{eq:nn}) represents the inverse reaction of pion production in nucleon-nucleon collisions, {\it i.e.}, 
$\mathrm{Im}\,a_{\pi\mathrm{D}}$ is also a measure of the strength of s-wave pion production at threshold. For that reason, pion-production 
cross sections were used to estimate the $\pi$D level width before it was experimentally accessible\,\cite{Hue75}. First 
attempts have been made to calculate the production strength rigorously also within the framework of $\chi$PT\,\cite{Len06,Len07} 
to be compared with the precise atomic data. 

The relative strength of the dominant channels true absorption and radiative capture (\ref{eq:nng}) was measured to be 
$S=\Gamma(\pi^-d\rightarrow nn)/\Gamma(\pi^-d\rightarrow nn\gamma)=2.83\pm\,0.04$\,\cite{Hig81}. The 
branching ratio of internal pair conversion (\ref{eq:nnee}) was determined to be 0.7\%\,\cite{Jos60}. Charge exchange 
(\ref{eq:nnpi0}) is parity forbidden from s states and, therefore, only odd partial waves contribute which results 
in a fraction as small as $(1.45\pm 0.19)\cdot 10^{-4}$\,\cite{Don77}. The corresponding partial $3p$ level width 
can be neglected here. Hence, the relative strength of true absorption to all other final states contributing to 
the $1s$ level width is found from the measured branching ratios to be 
\begin{eqnarray}
S'=\frac{\Gamma(nn)}{\Gamma(nn\gamma)+\Gamma(nne^+e^-)}=2.76\pm\,0.04\,,\label{eq:BR}
\end{eqnarray}
{\it i.e.}, about 2/3 of the hadronic width is related to the pion production/absorption process $NN\leftrightarrow\pi NN$. 
The exact relation between pion-production strength and $\mathrm{Im}\,a_{\pi\mathrm{D}}$ is derived in sect.\,\ref{sec:strong_interaction_width}. 

The aim of this experiment is to provide data on the $\pi$D hadronic shift and width at least at about the level 
of accuracy achieved in recent or envisaged in ongoing theoretical calculations. A precise value for the shift will 
provide, together with the forthcoming new precision data for $\pi$H\,\cite{PSI98,Got08}, improved constraints 
for the $\pi N$ scattering lengths $a^+$ and $a^-$. As a first result of this experiment, the s-wave pion 
production strength derived from the width has been outlined recently\,\cite{Str10}.

\section{Atomic cascade}\label{sec:cascade}

\begin{figure}[b]
\begin{center}
\resizebox{0.48\textwidth}{!}{\includegraphics{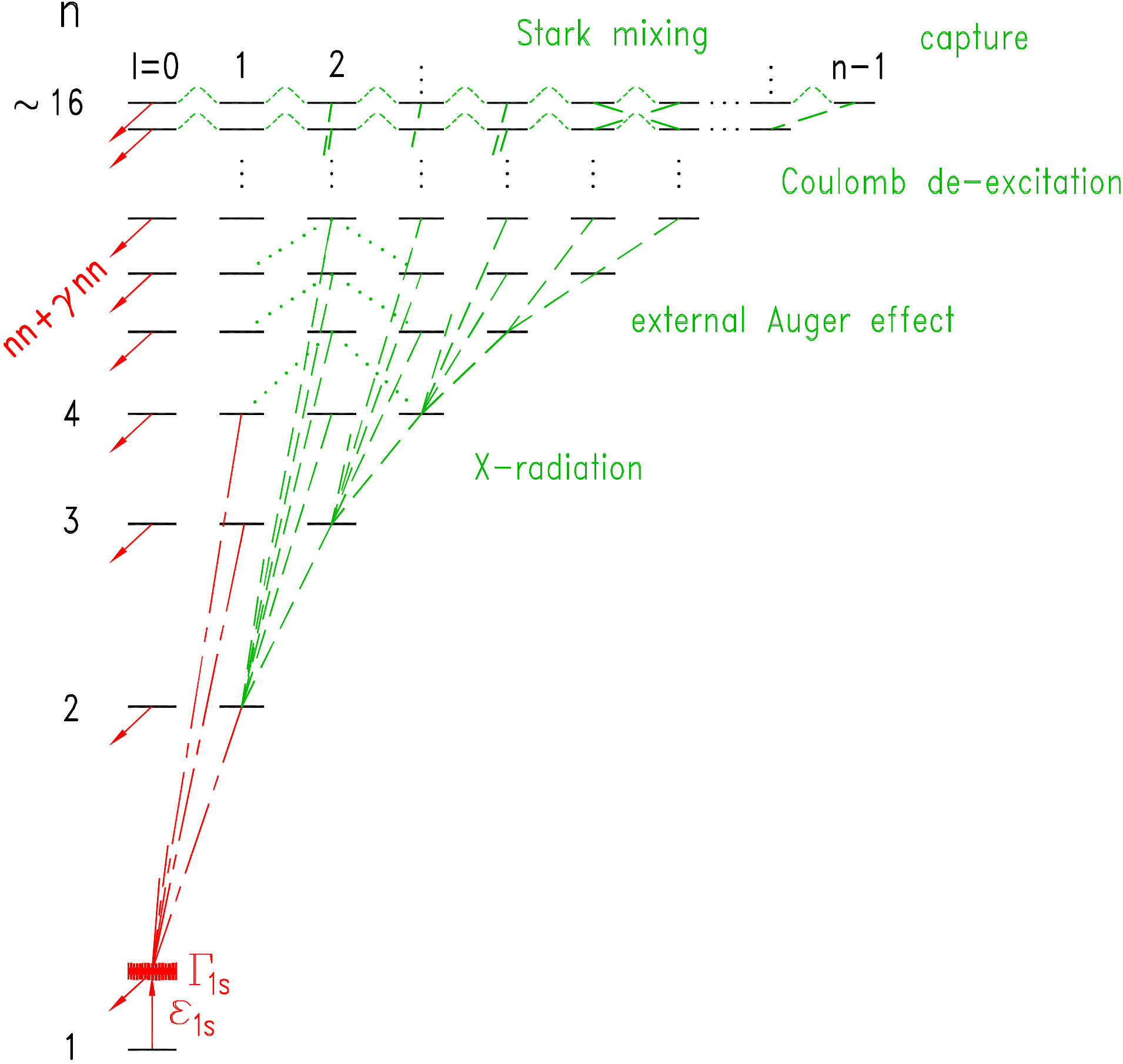}}
\caption{De-excitation cascade in pionic deuterium. The energy of the X-ray emitted in the $\pi$D$(3p-1s)$ transitions 
         is 3.075\,keV. Hadronic shift $\epsilon_{1s}$ and width $\Gamma_{1s}$ are of the order of 2.5 and 1\,eV, respectively. 
         The hadronic shift is defined as $\epsilon_{1s}\equiv \mathrm{E_{exp}-E_{QED}}$, {\it i.e.}, a negative  
         sign as is the case for $\pi$D corresponds to a repulsive interaction.}
\label{figure:fig1_piD_cascade}
\end{center}
\end{figure}

After pion capture in hydrogen isotopes, a quantum cascade starts from main quantum numbers 
$n\approx 16$\,\cite{Har90,Coh04} (Fig.\,\ref{figure:fig1_piD_cascade}). The upper and medium part of the atomic 
cascade is dominated by the collisional processes inducing Stark mixing, Coulomb de-excitation, and ionization 
of neighbouring atoms (external Auger effect). In the lower part, X-ray emission becomes more and more important. 
In light atoms, only electric-dipole transitions contribute preferring steps with maximal $\mathrm{\Delta}n$. The 
cascade time, estimated to be of the order of 0.1\,ns for densities around 10\,bar\,\cite{Bor80}, is much 
shorter than the pion's life time. In addition, molecular formation $\pi $D + D$_2\rightarrow [(dd\pi )d]ee$ 
must be considered. The influence of the above-mentioned processes is discussed in the following.

Radiative de-excitation is the only process where collisional effects on the line energy and width can be 
neglected. Doppler broadening from thermal motion is of the order of 10\,meV. Pressure broadening amounts at 
maximum to 0.2\,meV for the densities and temperatures used in this experiment. 

Stark mixing essentially determines the X-ray yields in exotic hydrogen. Because these systems are electrically 
neutral and small on the atomic scale, they penetrate surrounding atoms and, thus, experience a  
strong Coulomb field. Non-vanishing matrix elements $\langle nlm{\mid}\,{\bf E}{\mid}\,nl'm'\rangle$ in the presence 
of the electric field mix atomic states of the same principle quantum number {\em n} according to the selection 
rules $\mathrm{\Delta}l=\pm 1$ and $\mathrm{\Delta}m=0$\,\cite{Leo62}. An induced $s$ state, from where pions 
disappear by nuclear reactions, leads to a depletion of the X-ray cascade. As the Stark mixing rate is proportional to 
the number of collisions during the exotic atom's life time, it explains the strong decrease of the X-ray yields 
with target density\,\cite{Bor80}. The yield of the $\pi$D$(3p-1s)$ X-ray transition has been measured at target 
densities equivalent to 15\,bar and 40\,bar STP to be $(3.28\pm 0.43)$\% and $(1.72\pm 0.16)$\%, respectively\,\cite{Has95}. 
 
The natural line width of the atomic ground state transition $(3p-1s)$ of about 1\,eV is dominated by 
the life time of the $1s$ state. Nuclear reactions from the 3p level contribute with less than 1\,$\mu$eV 
and the $3p$-level radiative width amounts to 28\,$\mu$eV only. An estimate for the induced width from Stark mixing 
($3p\leftrightarrow 3s$) and external Auger effect at the $3p$ state, based on transition rates for $\pi$H given in 
ref.\,\cite{Jen02a}, yields also negligibly small contributions of $\leq 1\,\mu$eV at the target densities 
considered here. With a hadronic shift in $\pi$D being a factor of about 3 smaller than in $\pi$H, Stark mixing 
might be enhanced, but strong effects are excluded because similar K yields have been observed for hydrogen 
and deuterium\,\cite{Has95}.

However, significant broadenings of the X-ray line induced by other cascade processes must be considered, in particular in the 
case of Coulomb de-excitation\,\cite{BF78}. Here, the energy release during the transition $(n\rightarrow n')$ is converted 
into kinetic energy of the collision partners being the excited $\pi$D system and, predominantly, one atom of an 
D$_2$ molecule\,\cite{Jen02a,Jen02b,Jen02c}. Such transitions dominate de-excitation from the start of the quantum cascade down 
to $n\approx 10$ and have been found to contribute even at lowest densities\,\cite{Poh06}.

In lower-lying transitions, a significant energy gain occurs leading to a substantial Doppler broadening of 
subsequent X-ray transitions. The Doppler effect was directly observed in $\pi$H in time-of-flight spectra of 
monoenergetic neutrons from the charge exchange reaction at rest $\pi^-p\rightarrow \pi^0n$\,\cite{Czi63,Bad01} 
and in an additional broadening of the line width of the muonic hydrogen $(3p-1s)$ X-ray 
transition\,\cite{Cov09}. The broadening turned out to be a superposition of several Doppler contributions 
attributed to different de-ex\-citation steps $n\rightarrow n'$ of the initial $\mu^-p$ system. Only  
$\mathrm{\Delta}n$=1 transitions could be identified. A similar behaviour was found for pionic hydrogen. 
Here, a significant increase of the total line width is observed with decreasing $n$ of the initial state, 
which is attributed to the increasing energy gain of preceding Coulomb de-excitation steps\,\cite{Got08,Hen03}.

\begin{figure}[b]
\begin{center}
\resizebox{0.48\textwidth}{!}{\includegraphics{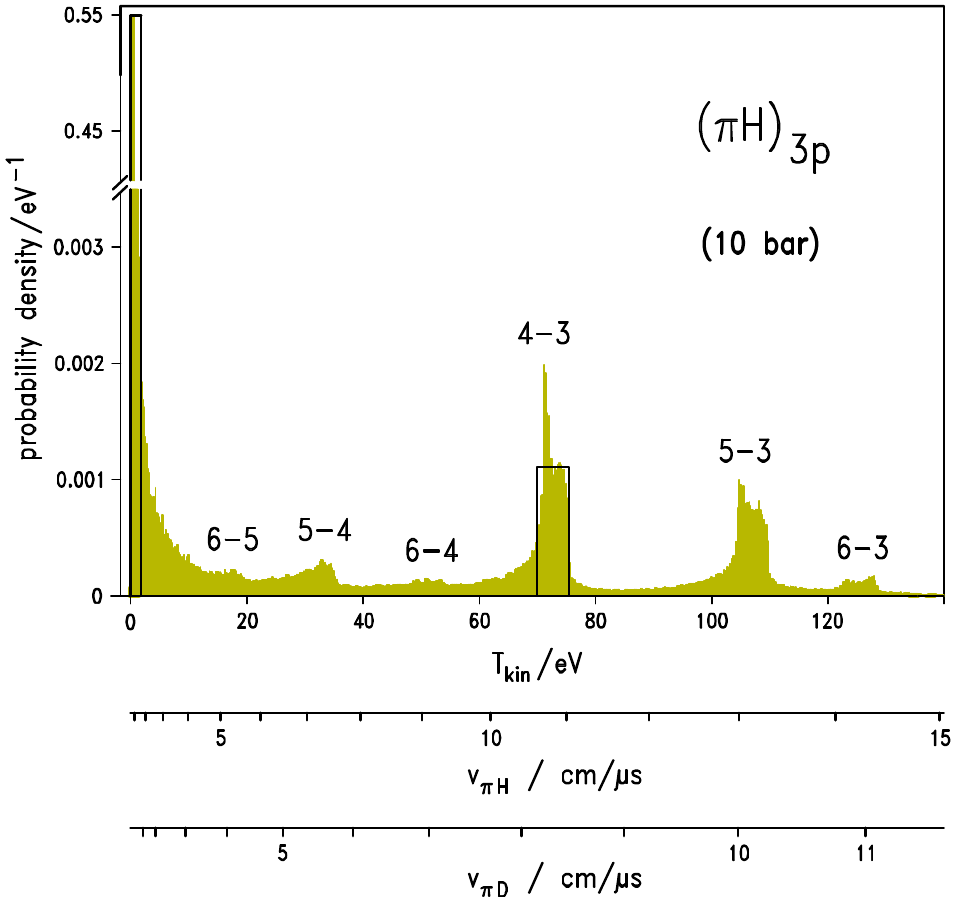}}
\caption{Prediction for the kinetic energy distribution of the $\pi$H atom at the instant of $(3p-1s)$ transition 
         for a density equivalent to 10\,bar gas pressure together with the velocity scale. The velocities for the 
         heavier $\pi$D system owing to the corresponding energy gains are given on the bottom. For the analysis 
         of this $\pi$D experiment, the energy distribution, after scaling the energies from the $\pi$H to the $\pi$D
         case, was tentatively approximated by two narrow intervals representing a low-energy component (0-2\,eV)  
         and a second one for the high-energy component stemming from the $(4-3)$ Coulomb de-excitation transition 
         (for details see sect.\,\ref{subsec:line_width}). 
         The low-energy component dominates the distribution (note the broken vertical scale).}
\label{figure:Tkin_ESCM}
\end{center}
\end{figure}

Acceleration due to Coulomb de-excitation is counteracted by elastic and inelastic scattering. This leads 
to a continuum of velocities below the ones well defined by a specific transition. Cascade calculations have 
been extended to include the development of the velocity dependence during de-excitation and, therefore, are 
able to predict kinetic energy distributions at the time of X-ray emission from a specific level (extended 
standard cascade model ESCM \cite{Jen02a,Jen02b,Jen02c}). At present, calculations have been performed only for 
$\pi$H\,\cite{Jen02a,Jen02b,Jen02c,PP06,JPP07,PP07}. Figure\,\ref{figure:Tkin_ESCM} shows such an ESCM prediction for the 
$3p$ state.

In the case of $\pi$D, the energies for the $\mathrm{\Delta}n=1$ Coulomb de-excitation transitions $(7-6)$, $(6-5)$, $(5-4)$, 
and $(4-3)$, are at 12, 20, 38, and 81\,eV, respectively.  
Though kinetic energies differ only little from the $\pi$H case, velocities are significantly smaller 
(Fig.\,\ref{figure:Tkin_ESCM}) possibly changing the balance between acceleration and slowing down.

In the analysis of a measurement of the $\mu$H$(3p-1s)$ line shape it turned out\,\cite{Cov09}, that first 
ESCM predictions are hardly able to describe the measured line width. Therefore, a model independent approach 
will be used to extract the relative strength of Doppler contributions directly from the X-ray data, which 
was applied successfully first in the neutron time-of-flight\,\cite{Bad01} and also in the $\mu$H$(3p-1s)$ 
analysis. Here, the continuous kinetic energy distribution is replaced by a few narrow intervals at the 
energies, where significant contributions are expected. Positions and intensities of such contributions are 
then found by comparing a Monte-Carlo generated line shape to the data by means of a $\chi^2$ analysis.
An approximation of the kinetic energy distribution by two narrow intervals, as used in the analysis of this  
experiment, is illustrated in fig.\,\ref{figure:Tkin_ESCM}. A detailed description is given in 
sect.\,\ref{subsec:line_width}. 

Auger transitions prefer de-excitation steps as small as possible, given that the energy gain exceeds the binding 
of the electron. Auger emission contributes mainly in the range $\mathrm{\Delta}n\approx 6-10$ with $\mathrm{\Delta}n=1$ 
transitions, but rapidly decreases for smaller $n$\,\cite{Bor80,Jen02b,Jen02c}. Because of the small recoil energies, 
Auger emission cannot contribute to the high energetic part of the kinetic energy distribution. 

From muon-catalysed fusion experiments it is known that during $\mu $H + H$_2$ collisions metastable hybrid molecules 
are formed like $[(pp\mu )p]ee$ \cite{Taq89,Jon99}. An analogous process is expected in $\pi $D + D$_2$ collisions\,\cite{Jon99}.  
Evidence that resonant molecular formation takes also place from excited states has been found in muonic hydrogen\,\cite{Poh09}. 

Such complex' are assumed to stabilise non radiatively by Auger emission. Possible X-ray transitions 
from weakly bound molecular states before stabilisation would falsify the value for the hadronic shift determined 
from the measured X-ray energy due to satellite lines shifted to lower energies, whereas Auger stabilised molecules 
emit X-rays of at least 30\,eV energy less than the undisturbed transition\,\cite{Kil04} and are easily 
resolved in this experiment. Radiative decay out of such molecular states has been discussed and its probability is 
predicted to increase with atomic mass\,\cite{Lin03,Kil04}. A weak evidence for an Auger stabilised transition is 
discussed in sect.\,\ref{subsec:energy}. 

As molecular formation is collision induced, the fraction of formed complex molecules and the X-ray rate should depend on 
the target density. Hence, a measurement of the $\pi$D$(3p-1s)$ X-ray energy was performed at different densities to identify 
such radiative contributions (see sect.\,\ref{subsec:energy_calibration}).

\section{Strong interaction}\label{strong_interaction}

In this section, the relations are introduced which will be used to extract the complex pion-deuteron 
scattering length $a_{\pi\mathrm{D}}$ (sect.\,\ref{subsec:scatteringlength}) and the threshold parameter for s-wave pion 
production (sect.\,\ref{subsec:alpha}) from the measured hadronic level shift and broadening.

\subsection{Scattering length}\label{sec:strong_interaction_shift}

The hadronic pion-nucleus scattering length $a_{\pi\mathrm{A}}$ is related in leading order (LO) to the $ns$ level shifts 
$\epsilon_{ns}$ and widths $\Gamma_{ns}$ by the Deser-Goldberger-Baumann-Thirring (DGBT) formula\,\cite{Des54}
\begin{eqnarray}
\epsilon_{ns}-i\,\frac{\Gamma_{ns}}{2} &=& -\frac{2\alpha^{3}\mu^{2}c^{4}}{\hbar c}\cdot Z^3\cdot\frac{a^{\mathrm{LO}}_{\pi\mathrm{A}}}{n^3}\label{eq:deser},
\end{eqnarray}
with $\mu$ being the reduced mass of the bound system of pion and nucleus $A(Z,N)$. In 
eqs.\,(\ref{eq:deser}) - (\ref{eq:true_im}) $\alpha$ denotes the fine structure constant.

Shift and width are measured using a Coulomb bound state. Therefore, the complex quantity $a^{\mathrm{LO}}_{\pi\mathrm{A}}$ 
as extracted from eq.\,(\ref{eq:deser}) is not identical with the pure hadronic scattering length 
$ a_{\pi\mathrm{A}}$. Usually, the Coulomb interaction is taken into account by Trueman's formula\,\cite{Tru61,Lam6970,Mit01} 
which expresses the correction by an expansion in the ratio scattering length-to-Bohr radius. In the case of 
hydrogen ($Z=1$), the complete expansion can be written up to order $O(\alpha^4)$ in a compact form\,\cite{Gas08,Lyu00} yielding for the 
$1s$ state of pionic deuterium\,\cite{Bar11} 
\begin{eqnarray}
\epsilon_{1s}-i\,\frac{\Gamma_{1s}}{2}&=& -\frac{2\alpha^{3}\mu^{2}c^{4}}{\hbar c}\,a_{\pi\mathrm{D}}\label{eq:moddeser}\\
                                      & &\cdot\,[1-\frac{2\alpha\mu c^2}{\hbar c}\,(\mathrm{ln}\,\alpha -1)\cdot a_{\pi\mathrm{D}}+\delta^{vac}_{\mathrm{D}}]\,.\nonumber
\end{eqnarray}
The term $\delta^{vac}_{\mathrm{D}}$=0.51\% accounts for the interference of vacu\-um polarisation and strong 
interaction\,\cite{Eir00}. Its uncertainty is assumed to be negligibly small compared to the experimental accuracy\,\cite{Gas08}.

In the case of $\pi\mathrm{D}$, the total correction to $a_{\pi\mathrm{D}}$ is only about 1\% (see eqs. (\ref{eq:Re_corr}) and (\ref{eq:Im_corr})). 
Hence, eq. (\ref{eq:moddeser}) is solved with sufficient accuracy by inserting the leading order result for 
$a^{\mathrm{LO}}_{\pi\mathrm{D}}$ as obtained from eq. (\ref{eq:deser}) and the influence of the experimental uncertainty must not be considered here. 

Real and imaginary part of the pure hadronic $\pi$D scattering length are then given by
\begin{eqnarray}
\mathrm{Re}\,a_{\pi\mathrm{D}}&=&-\frac{\hbar c}{2\alpha^{3}\mu^{2}c^{4}}\cdot \epsilon_{1s}\label{eq:true_re}\\
              & &\cdot\frac{1}{1-\frac{2\alpha\mu c^2}{\hbar c}\,(\mathrm{ln}\,\alpha -1)\cdot\frac{(\mathrm{Re}\,a^{\mathrm{LO}}_{\pi\mathrm{D}})^{2}-(\mathrm{Im}\,a^{\mathrm{LO}}_{\pi\mathrm{D}})^{2}}{\mathrm{Re}\,a^{\mathrm{LO}}_{\pi\mathrm{D}}}+\delta^{vac}_{\mathrm{D}}}\nonumber\\
\mathrm{Im}\,a_{\pi\mathrm{D}}&=&-\frac{\hbar c}{2\alpha^{3}\mu^{2}c^{4}}\cdot\frac{\Gamma_{1s}}{2}\label{eq:true_im}\\
              & &\cdot\frac{1}{1-\frac{2\alpha\mu c^2}{\hbar c}\,(\mathrm{ln}\,\alpha -1)\cdot 2\,\mathrm{Re}\,a^{\mathrm{LO}}_{\pi\mathrm{D}}+\delta^{vac}_{\mathrm{D}}}\,.\nonumber
\end{eqnarray}

\subsection{Threshold parameter $\alpha$ and pion production}\label{sec:strong_interaction_width}

The cross section for the reaction $pp\rightarrow~\pi^{+}d$ is parametrised at low energies by \cite{Ros54}
\begin{equation}
\sigma_{pp\rightarrow~\pi^{+}d}=\alpha C^{2}_{0}\eta + \beta C^{2}_{1}\eta^{3} + ...~,\label{eq:pid_pp}
\end{equation}
where $\eta =p^{*}_{\pi}/m_{\pi}$ is the reduced momentum of the pion in the $\pi d$ rest frame. Here, $\alpha$ 
denotes the threshold parameter representing pure s-wave pion production. Approaching threshold, $\eta\rightarrow 0$, 
higher partial waves ($\beta,\,...$) vanish. A value for $\alpha$ is obtained by fits to the quantity 
$\sigma_{pp\rightarrow~\pi^{+}d}/\eta $ and then extrapolated to zero energy. In particular at low energy, the 
correction factors $C_{i}$ which take into account the Coulomb interaction are an important source of uncertainty 
in the determination. For example, the leading order correction $C^{2}_{0}$ has been calculated to be as large as 30\% 
with the additional difficulty to obtain an accurate error estimate\,\cite{Rei69,Mac06}. 

A second experimental approach to $\alpha$ is to exploit the $\pi$D ground state broadening, where uncertainties stemming from 
Coulomb correction factors and normalisation of cross sections are avoided. In order to derive in a model independent way the relation 
between $\alpha$ and $\mathrm{Im}\,a_{\pi\mathrm{D}}$ as obtained from pionic deuterium, purely hadronic (non-measurable) cross 
sections $\tilde{\sigma}$ are introduced to circumvent the problem of diverging Coulomb cross section at threshold. 
In this case $C_i\equiv 1$, and $\alpha$ and $\beta$ are pure hadronic quantities calculable, {\it e.g.}, in the framework of 
$\chi$PT. The hadronic production cross section then reads 
\begin{equation}
\tilde{\sigma}_{pp\rightarrow~\pi^{+}d}=\alpha\eta + \beta\eta^{3} + ...~.\label{eq:pid_pp_had}
\end{equation}
In addition, the reaction $np\rightarrow~\pi^{0}d$ can be used because in the limit of charge independence 
$2\cdot\sigma_{np\rightarrow~\pi^{0}d}=\sigma_{pp\rightarrow~\pi^{+}d}$. Restricting to pion-deuteron $s$ waves, in both 
processes the same transition of a nucleon pair $^{3}P_{1}(I=1)\rightarrow\,^{3}S_{1}(I=0)$ occurs, where true 
pion absorption $\pi d\rightarrow nn$ induces the inverse reaction $^{3}S_{1}[^{\,3}D_{1}](I=0)\rightarrow\,^{3}P_{1}(I=1)$ 
on the deuteron's isospin 0 nucleon-nucleon pair.

Detailed balance relates pion production and absorption by 
\begin{eqnarray}
\tilde{\sigma}_{\pi^{+}d\rightarrow\,pp}& = &\frac{2}{3}\cdot \left(\frac{p^{*}_{p}}{p^{*}_{\pi}}\right)^2\cdot \,\tilde{\sigma}_{pp\rightarrow~\pi^{+}d}\label{eq:DB}
\end{eqnarray}
with $p^{*}_{p}$ and $p^{*}_{\pi}$ being final state centre-of-mass (CMS) momenta\,\cite{Bru51}. 
Neglecting Coulomb and isospin breaking corrections, charge symmetry requires for the transitions 
$\pi^{-}d~\rightarrow~nn$ and $\pi^{+}d~\rightarrow~pp$
\begin{equation}
\mid M_{\pi^-d\rightarrow nn}\mid =\mid M_{\pi^+d\rightarrow pp}\mid.\label{eq:mirror}
\end{equation}
Isospin breaking effects are expected to be at most 1-2\%\,\cite{Barpc,Fil09}.

A small difference in the transition rate results from the slightly larger phase space of the $\pi^+d\rightarrow pp$ 
reaction. 
\begin{eqnarray}
\frac{\tilde{\sigma}_{\pi^-d\rightarrow nn}}{\tilde{\sigma}_{\pi^+d\rightarrow pp}}
    =\frac{p^{*}_{n}}{p^{*}_{p}}\,
    =\sqrt{\frac{\lambda(s,m^2_n,m^2_n)}{\lambda(s,m^2_p,m^2_p)}}
    =0.982\,,\label{eq:psp}
\end{eqnarray}
The CMS momenta of proton and neutron, expressed in invariant variables using the triangle function $\lambda$, 
are given by $p^*_{p,n}=\lambda^{1/2}(s,m^2_{p,n},m^2_{p,n})/2\sqrt{s}$ \cite{Byc73}. The value corresponds to 
$\pi^{\pm}d$ at threshold, where the total CMS energy squared is $s=(m_d+m_{\pi^{\pm}})^2$. The atomic binding 
energy of the $\pi^-$D system is neglected.

Combining optical theorem, charge invariance, detailed balance and inserting the parametrisation 
of the $pp\rightarrow \pi^+d$ cross section (\ref{eq:pid_pp_had}), the imaginary part of the $\pi^{-}d\rightarrow nn$ 
scattering length reads in terms of the $\pi^{+}$ production threshold parameter $\alpha$  
\begin{eqnarray}
\mathrm{Im}\,a_{\pi^- d\rightarrow nn}&=&\frac{p^*_{\pi}}{4\pi }\cdot \tilde{\sigma}_{\pi^-d\rightarrow nn}\nonumber\\ 
             &=&\frac{p^*_{\pi}}{4\pi }\cdot \tilde{\sigma}_{\pi^+d\rightarrow pp}\cdot\left(\frac{p^*_{n}}{p^*_{p}}\right)\nonumber\\ 
             &=&\frac{p^*_{\pi}}{4\pi }\cdot \frac{2}{3}\cdot\left(\frac{p^*_{p}}{p^*_\pi}\right)^2\cdot \tilde{\sigma}_{pp\rightarrow \pi^+d}\cdot\left(\frac{p^*_{n}}{p^*_{p}}\right)\nonumber\\
             &=&\frac{1}{6\pi}\cdot \frac{(p^*_{p}\cdot p^*_{n})}{m_{\pi}}\cdot\alpha\,.\label{eq:Ima_alpha} 
\end{eqnarray}

To relate $\mathrm{Im}\,a_{\pi^- d\rightarrow nn}$ given by pion production to the imaginary part of the pion-deuteron scattering length \linebreak
$\mathrm{Im}\,a_{\pi D}$ as obtained from the atomic system $\pi$D, a correction must be applied for the 
non true absorption channels. Taking into account the relative strength $S'$ of true absorption to other s-state 
processes (\ref{eq:BR}) and exploiting $\lambda(x,y,y)=x(x-4y)$\,\cite{Byc73}, one obtains
\begin{eqnarray}
\mathrm{Im}\,a_{\pi\mathrm{D}}&=&(1+\frac{1}{S'})\cdot \mathrm{Im}\,a_{\pi^- d\rightarrow nn}\nonumber\\
              &=&(1+\frac{1}{S'})\cdot\,\frac{\sqrt{(s-4m^2_p)(s-4m^2_n)}}{24\pi\,m_{\pi}}\cdot\alpha\nonumber\\
              &=&2.48\cdot 10^{-5}\,m^{-1}_{\pi}\,\mu b^{-1}\,\cdot\,\alpha\,.\label{eq:Ima_nr}
\end{eqnarray}

\section{Experimental setup}\label{sec:exp}

\begin{figure}[b]
\begin{center}
\resizebox{0.485\textwidth}{!}{\includegraphics{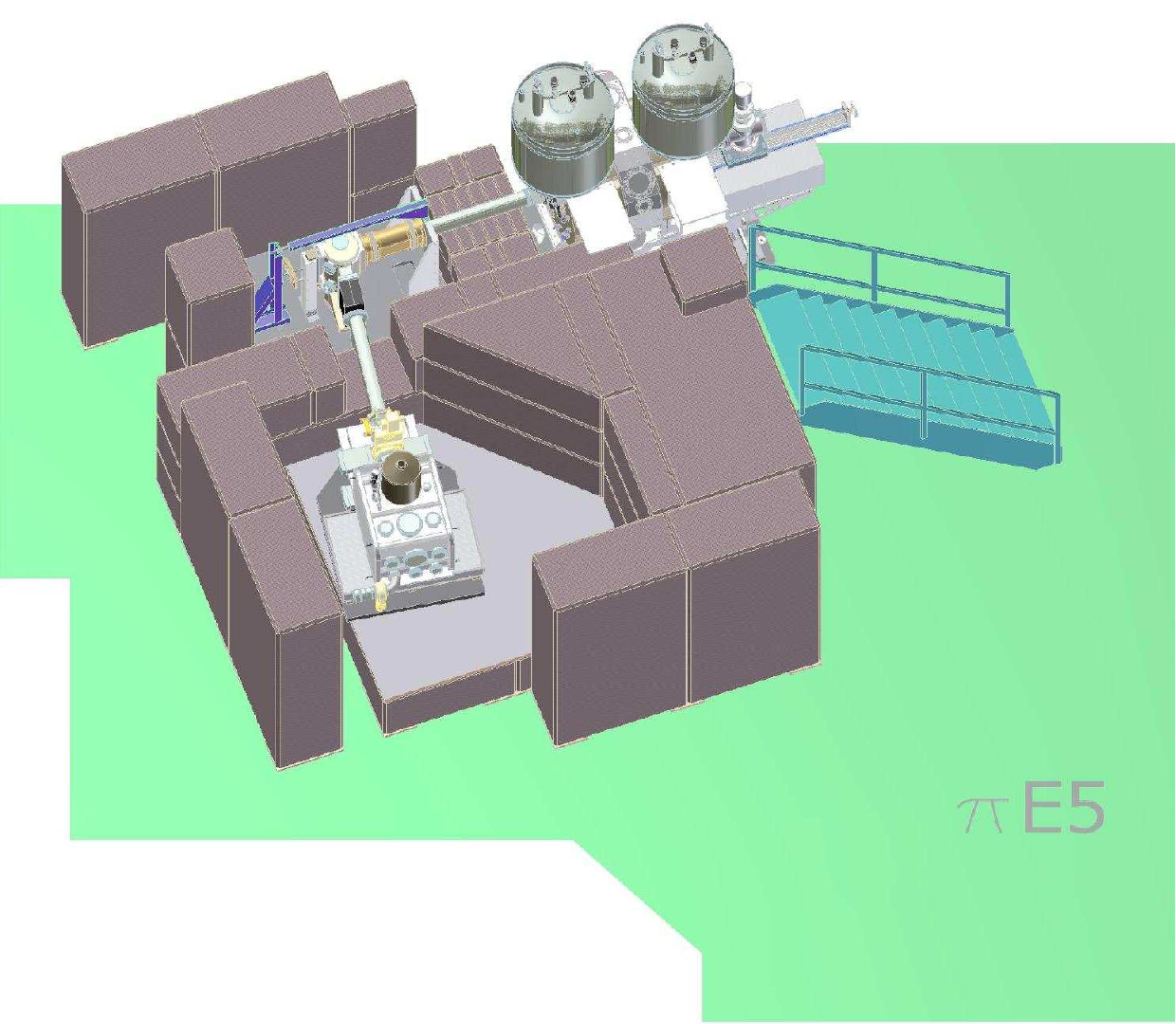}}
\caption{Setup of the $\pi$D experiment in the $\pi$E5 area at the Paul Scherrer Insitut (PSI). The roof 
         of the concrete shielding is omitted to show the vacuum system connecting the cyclotron trap (upper 
         right), crystal chamber (upper left) and the cryostat of the X-ray detector (bottom left). 
         Details of the mechanical setup are shown in figs.\,\ref{figure:cyc-cry} and \ref{figure:cry-det}.}
\label{figure:setup_piD}
\end{center}
\end{figure}

The experiment was performed at the $\pi$E5 channel of the proton accelerator at PSI, which provides a 
low-energy pion beam with intensities of up to a few $10^{8}$/s (Fig.\,\ref{figure:setup_piD}). Pions of 112\,MeV/c 
were injected into the cyclotron trap II\,\cite{Sim88,Sim93} and decelerated by using a set of degraders optimised 
to the number of pion stops in a cylindrical cryogenic target cell of 22\,cm length and 5\,cm in diameter placed 
in the centre of the trap. Its magnetic field, perpendicular to the beam direction, guides the particles after a few turns 
towards the target.

The cell was filled with deuterium gas cooled by means of a cold finger to adjust the target density by temperature. Data 
were taken at three different conditions, which are equivalent to gas densities of 3.3, 10, and 17.5\,bar at a temperature 
of 20$^{\circ}$C. The corresponding target parameters may be found in table\,\ref{table:exp}. About 0.5\% of the 
incoming pions are stopped per bar equivalent pressure of D$_{2}$. X-radiation could exit the target cell axially through 
a 5\,$\mu$m thick mylar$^{\textregistered}$ window. The window foil is supported by 1\,mm thick horizontal aluminum bars 
constructed with a 6\,mm spacing.

X-rays from the $\pi$D$(3p-1s)$ transition were measured with a Johann-type Bragg spectrometer equipped with a 
spherically bent Si crystal cut along the (111) plane and having a large radius of curvature of $R=2982.2\pm 0.3$\,mm 
in order to minimise aberrations. 
The crystal plate was attached by molecular forces to a glass lens polished to optical quality, which defines the 
curvature. A possible miscut and its orientation were determined in a dedicated measurement\,\cite{Cov08}. 
The reflecting area of 10\,cm in diameter was restricted by a circular aperture to 95\,mm in diameter 
to avoid edge effects and to 60\,mm horizontally in order to keep the Johann broadening small\,\cite{Egg65,Zsc82}. 

Such a spectrometer is able to measure simultaneously an energy interval according to the width of the 
X-ray source when using a correspondingly extended X-ray detector. Being pixel detectors, charge-coupled devices \linebreak  
(CCDs) are ideal detectors for X-rays in the few keV range because they combine an intrinsic position resolution 
with the good energy resolution of semiconductor detectors. In this setup an array of $3\times 2$ CCDs was 
used covering in total 72\,mm in height and 48\,mm in width\,\cite{Nel02}. 

\begin{figure}[b]
\begin{center}
\resizebox{0.485\textwidth}{!}{\includegraphics{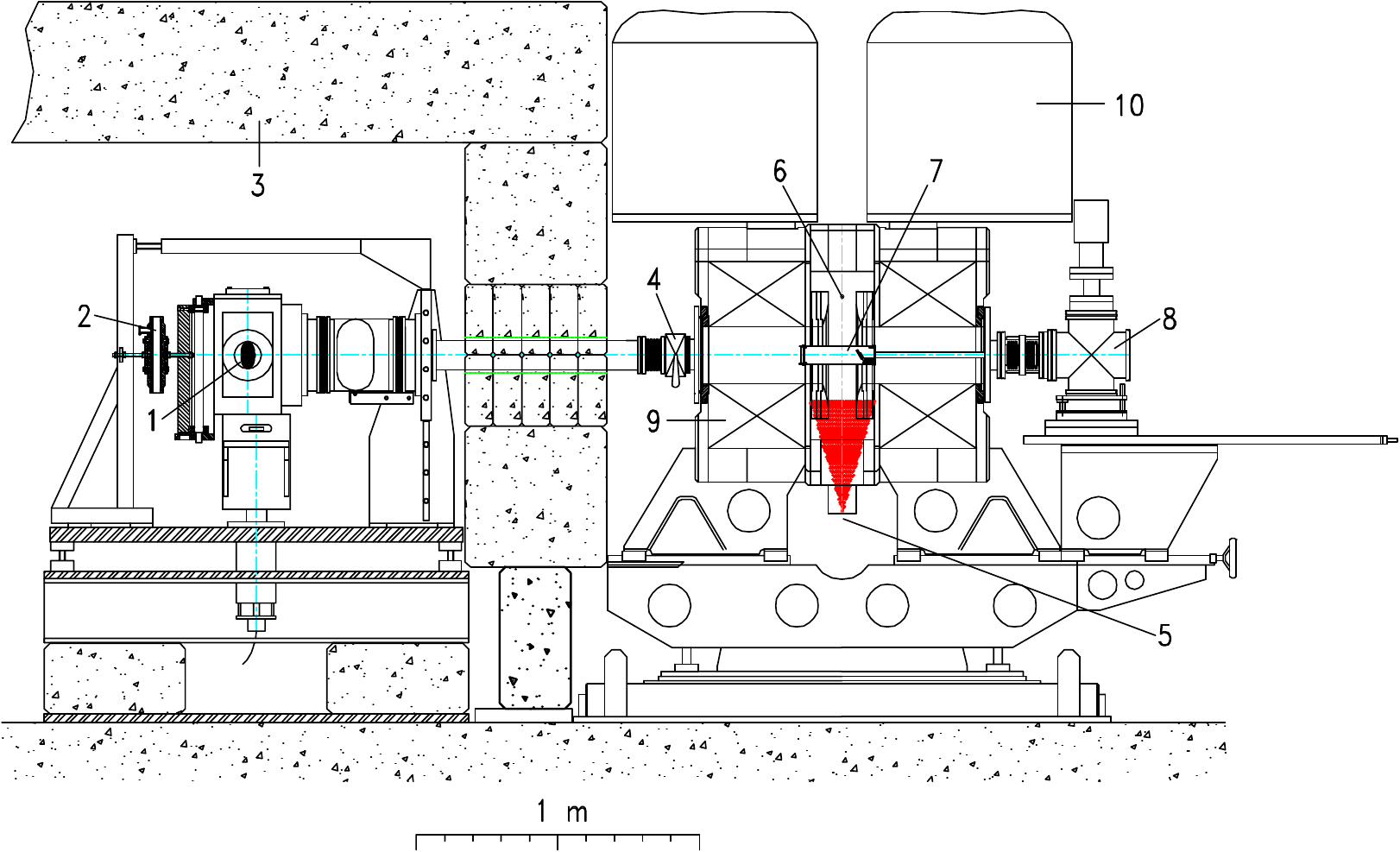}}
\caption  {Sectional drawing of the connection of the cyclotron trap to the crystal chamber of the spectrometer. 
           The symmetry axis of the magnetic field coincides with the line center of the trap to center of the Bragg crystal.
           1: Bragg crystal, 2: traction relaxation, 3: concrete shielding, 4: gate valve, 5: X-ray tube (emission cone indicated), 
           6: position of pion beam without B field, 7: cryogenic target cell including calibration target (right end), 
           8: target cell support and cryogenic generator, 9: magnet coils of cyclotron trap, 10: LHe dewars.}
\label{figure:cyc-cry}
\end{center}
\end{figure}

Monte Carlo studies show that 
about 2/3 of the intensity of the reflection is covered by the height of the CCD array. The solid angle of the 
crystal with respect to the target is $\approx 6\cdot 10^{-5}$ and the target fraction accepted by the crystal's 
angular width is $\approx 5\cdot 10^{-3}$. Including window absorption and assuming a peak reflectivity of 40\% 
for the Bragg crystal, the overall efficiency of the spectrometer results in about $10^{-7}$.

The CCDs are realised in open electrode architecture to minimise X-ray absorption in the surface structures and 
a depletion depth of about 30\,$\mu$m yields a quantum efficiency of about 80\% in the 3-4\,keV range. The pixel 
size of 40\,$\mu$m provides a sufficient two-dimensional position resolution to measure precisely the shape of the 
diffraction image. The relative orientation of the individual CCDs as well as the pixel size at the operating 
temperature of $-100^{\circ}$C was obtained by means of an optical measurement using a nanometric grid\,\cite{Ind06}. 
The detector surface was oriented perpendicular to the direction crystal -- detector. 
 
\setlength{\tabcolsep}{0.6mm}
\begin{table}
 \caption[Setup data PSI]
   {Data to adjust the spectrometer in Johann setup. The Bragg angle $\Theta_{\mathrm{B}}$ is calculated using 
    a lattice distance $d$ of 2d$_{111}$=(0.62712016\,$\pm$\,0.00000001)\,nm \cite{Bas94}. $n$ denotes the order of reflection. 
    For the $\pi$D$(3p-1s)$ energy, here the experimental result of ref.\,\cite{Cha9597} is given. 
    $y_{\mathrm{CD}}=R\cdot sin\,(\Theta_{\mathrm{B}}+\mathrm{\Delta}\Theta_{\mathrm{IRS}})$ are calculated focal lengths taking into account the 
    index of refraction shift $\mathrm{\Delta}\Theta_{\mathrm{IRS}}$. The Ga K$\alpha$ energies are taken from the compilation 
    of ref.\,\cite{Des03}.}
  \label{table:setpidga}
\begin{tabular}{llcccccc}
 \hline \\[-3mm]
                 & ~~~~~~~energy          & $n$& $\Theta_{\mathrm{B}}$ & $\mathrm{\Delta}\Theta_{\mathrm{IRS}}$ &$y_{\mathrm{CD}}$ \\
                 & ~~~~~~~\,/\,eV         &    &                    &                      & /\,mm   \\[1.0mm]
 \hline\\ [-3mm]
 Ga K$\alpha_1$  & 9251.674\,$\pm$\,0.066 & 3 & $39^{\circ}52'22.3"$&  2.4"                & 1912.42  \\[1.0mm]
 $\pi$D$(3p-1s)$ & 3075.52~\,\,$\pm$\,0.70   & 1 & $40^{\circ}0'11.7"$ &  21.8"               & 1917.84  \\[1.0mm]
 Ga K$\alpha_2$  & 9224.484\,$\pm$\,0.027 & 3 & $40^{\circ}0'44.1"$ &  2.4"                & 1917.98  \\[1.0mm]
  \hline
  \end{tabular}
\end{table}  

In the Johann setup, the energy calibration must be provided by a reference line of known energy. The best 
angular matching for the $\pi$D$(3p-1s)$ measurement was found with the gallium K$\alpha$ fluorescence lines. 
Here, the Ga K$\alpha_2$ transition was chosen because of its smaller experimental error (Table\,\ref{table:setpidga}). 
The energy of the measurement line is obtained by the angular distance to the reference line, which is calculated 
from the position offset of the two reflections on the detector and the distance crystal -- detector. 

The fluorescence target was made of a $25\times 20$\,mm$^2$ GaAs plate mounted in the rear part of the gas cell outside 
the pion stop volume and 82\,mm off the centre of the cyclotron trap away from the crystal (Fig.\,\ref{figure:cyc-cry}). 
X-rays were excited by means of an X-ray tube mounted on a window of the cyclotron trap chamber below the 
gas cell. 

\begin{figure}[b]
\begin{center}
\resizebox{0.485\textwidth}{!}{\includegraphics{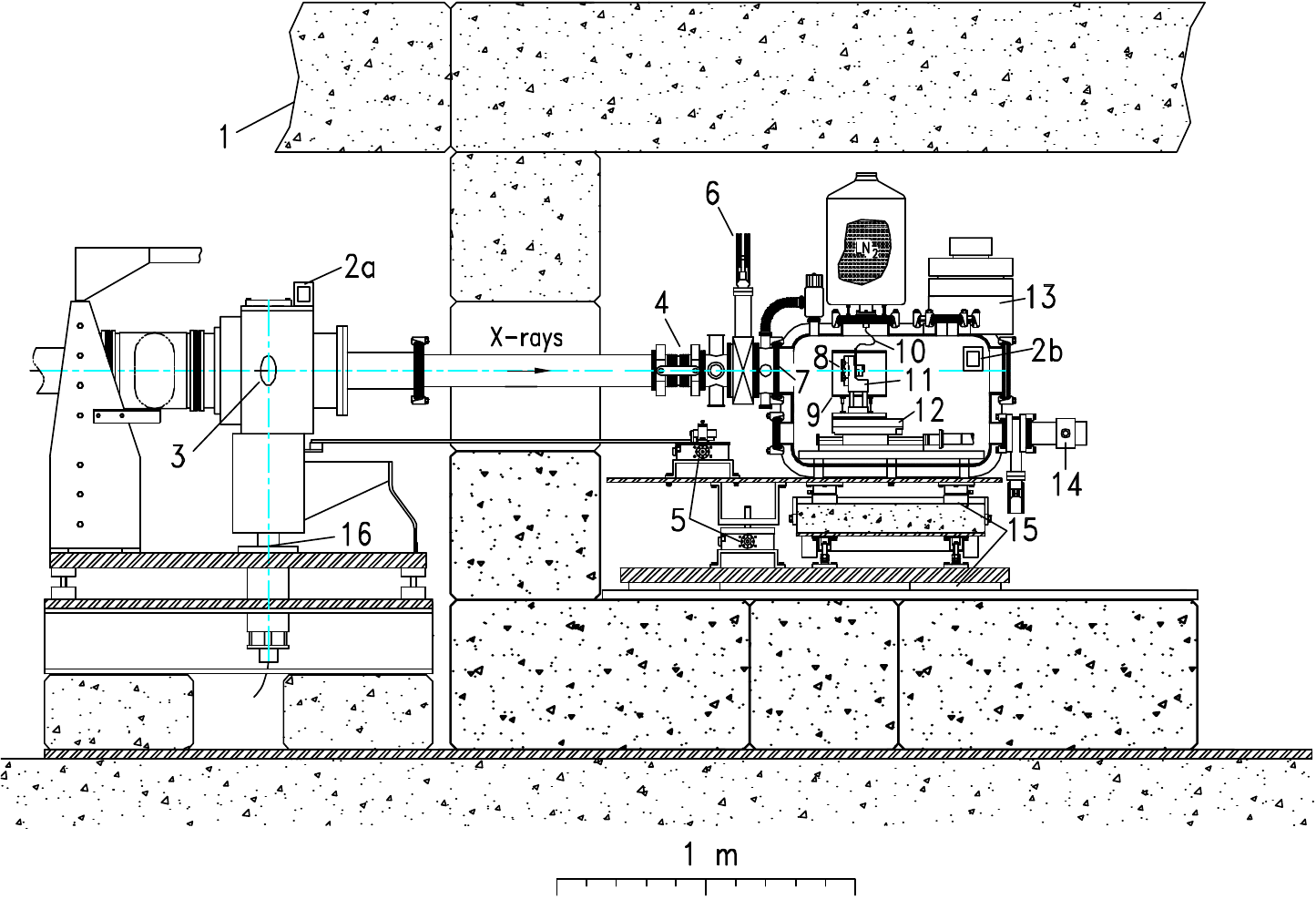}}
\caption {Sectional drawing of the connection of the crystal chamber to the X-ray detector. 
          1: concrete shielding, 2a and 2b: inclination sensors, 3: Bragg crystal, 4: compensator, 
          5: linear tables for crystal and detector adjustment, 
          6: gate valve, 7: 5~$\mu $m Mylar window, 8: CCD array, 9: cold trap, 10: copper braid, 11: cold finger, 
          12: translation table for focal length adjustment, 13: detector readout electronics, 14: turbomolecular pump, 
          15: air cushion support, 16: crystal chamber support.}
\label{figure:cry-det}
\end{center}
\end{figure} 

In order to exclude the distortion of the $\pi$D$(3p-1s)$ line shape from beam-induced excitation of Ga X-rays, 
data were taken using H$_{2}$ gas at an equivalent density of 10\,bar. Using a hydrogen filling guarantees similar 
stopping conditions for the pion beam. No Ga X-rays could be identified during a measuring time corresponding 
to about 15\% of the one used for D$_{2}$ at the same density. Other sources of background are discussed in 
sect.\,\ref{sec:analysis}.

The distance from the centre of the crystal to the centre of the cyclotron trap was 2100\,mm, about 10\% outside 
the Rowland circle given by the focal condition $R\cdot sin\Theta_{\mathrm{B}}$. The advantage of placing the X-ray source 
somewhat off the focal position is an averaging over non uniformities of the target. Both GaAs and D$_{2}$ target 
were extended enough that no cuts in the tails of the reflection occur. The distance crystal -- detector (Fig.\,\ref{figure:cry-det}), chosen 
to be at the assumed $\pi$D focal length, was found to be y$_{\mathrm{CD}}=1918.1\,\pm\,0.5$\,mm from a survey measurement.

Alternating measurements of the Ga fluorescence radiation and the $\pi$D line were performed at least once per day. 
The Ga fluorescence X-rays were in addition used to monitor the stability of the line position. 

The spectrometer response was measured at the energies 3104, 2765, and 2430\,eV using the narrow M1 X-ray lines 
from helium-like argon, chlorine, and sulphur produced in a dedicated electron-cyclotron ion resonance trap (ECRIT) 
\cite{Ana05,Tra07,Covth}. It turned out that the 
resolution function at a given energy can be built up from the ideal response calculated from the dynamical theory of 
diffraction for a perfect flat crystal (intrinsic resolution) convoluted with the geometrical imaging at the measurement 
position by means of a Monte-Carlo ray-tracing code and by folding in an additional Gaussian contribution. The 
intrinsic resolution is calculated here with the code XOP\,\cite{San98}. The Gaussian models possible imperfections 
of the crystal material and mounting and was found to be sufficiently precise, that no difference is visible by eye 
between data and Monte-Carlo simulations. 

The value for the Gaussian width at the energy of the $\pi$D$(3p-1s)$ 
transition energy of 3075\,eV was found from the fit to the results for argon, chlorine, and sulphur to be 
($122\pm$8)\,meV. The narrow structure in fig.\,\ref{figure:GaKa_piD10bar} - middle shows the Monte-Carlo generated 
response for the setup of the $\pi$D experiment. The total width of this resolution function is (436$\,\pm$\,3)\,meV 
(FWHM) and close to the theoretical limit of 403\,meV---calculated by means of XOP---for the intrinsic resolution 
of a silicon crystal cut along the (111) plane. 

Details on the experimental setup and analysis may be found elsewhere\,\cite{Str09}.

\section{Analysis}\label{sec:analysis}

Raw data of the X-ray detector consist of the digitized charge contents and a position index of the pixel. 
The granularity of the CCDs allows for efficient background rejection by means of pattern recognition (cluster 
analysis). Photoelectrons from few keV X-ray conversion are stopped within a few micrometer only and, therefore,
deposit charge in one or two pixels with one common boundary. Beam induced background, mainly high energetic 
photons from neutrons produced in pion absorption and captured in surrounding nuclei, lead to larger structures.
Together with a massive concrete shielding (Fig.\,\ref{figure:setup_piD}) such events are highly suppressed. 
Defect pixels are masked by software. 

\begin{figure}[t]
\begin{center}
\resizebox{0.45\textwidth}{!}{\includegraphics{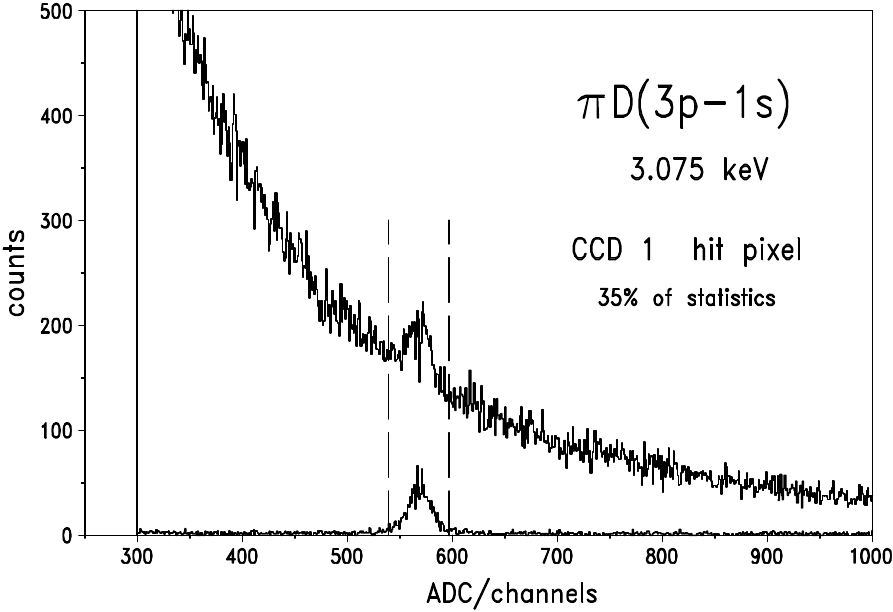}}
\caption{Spectrum of the collected (digitized) charge in one CCD before and after cluster analysis of one of the 
         3 CCDs covered by the $\pi$D$(3p-1s)$ reflection. Data are from the measurement at 10\,bar equivalent density
         and correspond to 35\% of the $\pi$D  statistics collected with CCD\,1. The  
         peak around ADC channel 570 is due to the transition energy of 3.075\,keV. Vertical dashed lines indicate the 
         applied "energy cut" of $\pm\,2.5\sigma$ allowing for further background reduction.}
\label{figure:ADC_CCD}
\end{center}
\end{figure}

At first, the cluster analysis is performed. As expected only single ($\approx$\,75\%) or two pixel events 
($\approx$\,25\%) contribute. The spectra of the collected charge cleaned in this way 
show a pronounced peak originating from the 3\,keV $\pi$D X-rays (Fig.\,\ref{figure:ADC_CCD}). For each CCD an individual 
energy calibration was performed by using the $\pi$D$(3p-1s)$ line, necessary because of the different gain and noise 
behaviour of the various devices. The second calibration point is zero because the noise peak is suppressed 
on-line during data acquisition which reduces the amount of data to be recorded substantially. The energy 
resolution in terms of charge is determined by means of a Gaussian fit and found to be between 170 and 
300\,eV at 3\,keV. 

\begin{figure}[h]
\begin{center}
\resizebox{0.45\textwidth}{!}{\includegraphics{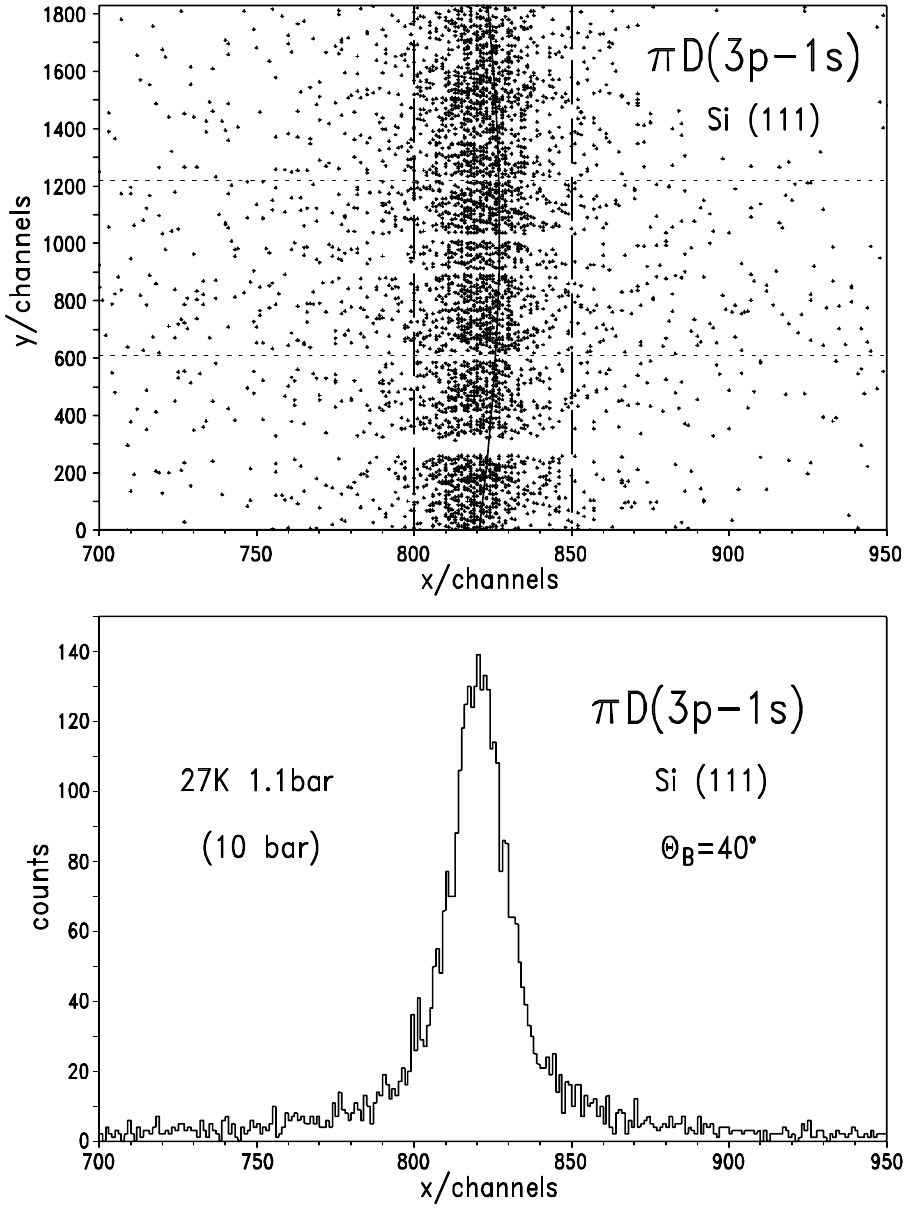}}
\caption{Top -- Scatter plot of the $\pi$D$(3p-1s)$ reflection measured at a density equivalent to 10\,bar (NTP). 
         Its curvature is obtained by a parabola fit. Bottom -- projection to the axis of dispersion after curvature 
         correction. One channel (pixel) in the direction of dispersion ($x$) corresponds in first order reflection 
         to 76.402$\pm$0.001\,meV.}
\label{figure:scat_piDline}
\end{center}
\end{figure}

The good resolution in energy ({\it i.e.}, in collected charge) by CCDs allows to apply a narrow "energy cut" in the 
ADC spectra (Fig.\,\ref{figure:ADC_CCD}). Thus an additional and significant background reduction is achieved in the 
two dimensional position spectra (Fig.\,\ref{figure:scat_piDline} -- top). Assuming a Gaussian shape for the energy 
resolution and setting various windows ranging from $\pm\,1\sigma$ to $\pm\,4\sigma$, the influence was studied of 
the peak-to-background ratio on the result for the hadronic broadening and found to be marginal. The minimum statistical 
error is achieved for a cut of $\pm\,2.5\sigma$. The X-ray count rates at the equivalent densities of 3.3, 10, and 
17.5\,bar are given in table\,\ref{table:exp}. 

The hit pattern of the X-rays on the CCD surface shows a curvature originating from the imaging properties of 
the crystal spectrometer. The curvature is corrected by means of a parabola fit before projecting 
onto the axis of dispersion, which is equivalent to an energy axis (Fig.\,\ref{figure:scat_piDline}).

The penetration depths of the $\pi$D$(3p-1s)$ and the Ga X-rays differ significantly (5 and 105\,$\mu$m for 3.075 
and 9.224\,keV, respectively\,\cite{Vei73}). Ga K X-rays also convert at or beyond the boundary of the depletion 
region of the CCD, where charge diffusion 
is already significant. Therefore, in this case compact clusters up to size 9 were accepted. As the Ga calibration 
measurements were performed without pion beam in the experimental area, no background events occur and, hence, 
no suppression algorithm is necessary to clean the Ga spectra. For one Ga calibration measurement, typically 
4000 Ga K$\alpha_{2}$ events were recorded. The total line width was found to be $(2.9\pm 0.1)$\,eV, which is 
two standard deviations above the natural line width of $(2.66\pm 0.13)$\,eV as given in ref.\,\cite{Kra79}.

\setlength{\tabcolsep}{0.35mm}
\begin{table}[t]
  \caption{Experiment parameters. Rates (counts/Cb) are meant for a detector width of 14\,mm and normalised 
           to the integrated accelerator proton current (typically 2\,mA) for "energy cut" of $\pm\,2.5\sigma$. 
           The equivalent density of the target gas is expressed as pressure value at a temperature of 
           20$^{\circ}$C (NTP). The number of Ga calibration runs for each density is indicated in the last column. 
           The backgound rate is density independent and scales linearly with the width of the "energy cut". 
           The background given here stems from a detector area of 17.28\,cm$^{2}$ corresponding to 3 CCDs.}
 \label{table:exp}
\begin{tabular}{cccccc}
 \hline\\[-3mm]
  D$_2$ density &\multicolumn{2}{c}{target parameters}&\multicolumn{2}{c}{counts}            &no. of Ga\\
  (equivalent)  & T            & pressure             & \multicolumn{2}{c}{$\pi$D$(3p-1s)$}  &calibrations     \\
  / bar         &   /\,K       & /\,bar               & total         &/\,Cb                 & \\[1mm]
 \hline\\[-3mm]
   ~\,3.3\,$\pm$\,0.1 &   33      &   0.51               &1448$\pm$49    &1.98$\pm$0.07      & 10  \\
  10.0\,$\pm$\,0.3 &      27      &   1.09               &4010$\pm$74    &6.40$\pm$0.12      &  8  \\ 
  17.5\,$\pm$\,0.4 &      25      &   1.36               &4877$\pm$80    &7.35$\pm$0.12      &  7  \\[1mm] 
  \hline\\[-2mm]
  background       &              &                      &               &1.64$\pm$0.11      &    \\[1mm] 
  \hline
  \end{tabular}
\end{table}

\subsection{Energy calibration}
\label{subsec:energy_calibration} 
The tabulated values for the Ga K$\alpha$ energies\,\cite{Des03} were obtained using the compound GaAs\,\cite{Moo07} 
-- the same material used in this experiment, {\it i.e.}, a possible chemical shift is irrelevant for this calibration. 
The position of the Ga K$\alpha_2$ line was determined applying a single Voigt profile in the fit, which is also the 
procedure used for the tabulated values\,\cite{Moo07}. The matching of the angular positions owing to the Ga K$\alpha_2$ 
and $\pi$D$(3p-1s)$ energies is evident from fig.\,\ref{figure:GaKa_piD10bar}. 

The $\pi$D line was modeled both with a Voigt profile, where Doppler broadening and response function together are 
approximated by a single Gaussian, and the true response as determined from the ECRIT data convoluted with the 
imaging properties by means of a ray tracing MC code and including the Doppler contributions from 
Coulomb de-excitation. Both methods yield the same position value within a few hundreds of one CCD pixel.

Various corrections to the measured line positions must be applied. Significant are, besides the index of 
refraction shift, bending and penetration correction, a shift of the gallium line due to the non-uniform 
illumination of the fluorescence target, and mechanical shifts from temperature changes from the 
nitrogen filling of the detector. These corrections are discussed in the following and listed together with less 
significant corrections in table\,\ref{table:en_corr}.

\begin{itemize}
\item {\bf Index of refraction.} 
The Ga calibration line and the $\pi$D$(3p-1s)$ transition were measured in third and first order diffraction 
(table\,\ref{table:setpidga}). As the index of refraction strongly depends on the X-ray wave length a 
significant correction must be applied in order to obtain the right Bragg angle difference. The index of refraction 
$\mathrm{\Delta}\Theta_{\mathrm{IRS}}$ is calculated here from the atomic scattering factors as used by the 
code XOP\,\cite{San98}. The accuracy of such amplitudes is claimed to be of the order of 1\% for 3 and 9\,keV, 
respectively,  as are the angular corrections derived from them. The amplitudes of various sets vary by less than 
0.5\%\,\cite{Hen93,Cha95}.
\item {\bf Crystal bending.} 
Crystal bending leads to a depth dependence of the lattice spacing $d$, where close to the surface the maximum 
lattice distance is assumed. Because of the difference in energy and accordingly the different mean penetration 
depth of the X-rays, the average lattice spacing and, consequently, the diffraction angle is different. The change 
of $d$ depends on the elastic properties of the nonisotropic crystal material and is calculated following the 
approach of ref.\,\cite{Chu96} using the value $\nu'=0.182$ for Poisson's ratio together with an expression for 
the angular correction as given by ref.\,\cite{Cem92}. 

\begin{figure}[t]
\begin{center}
\resizebox{0.42\textwidth}{!}{\includegraphics{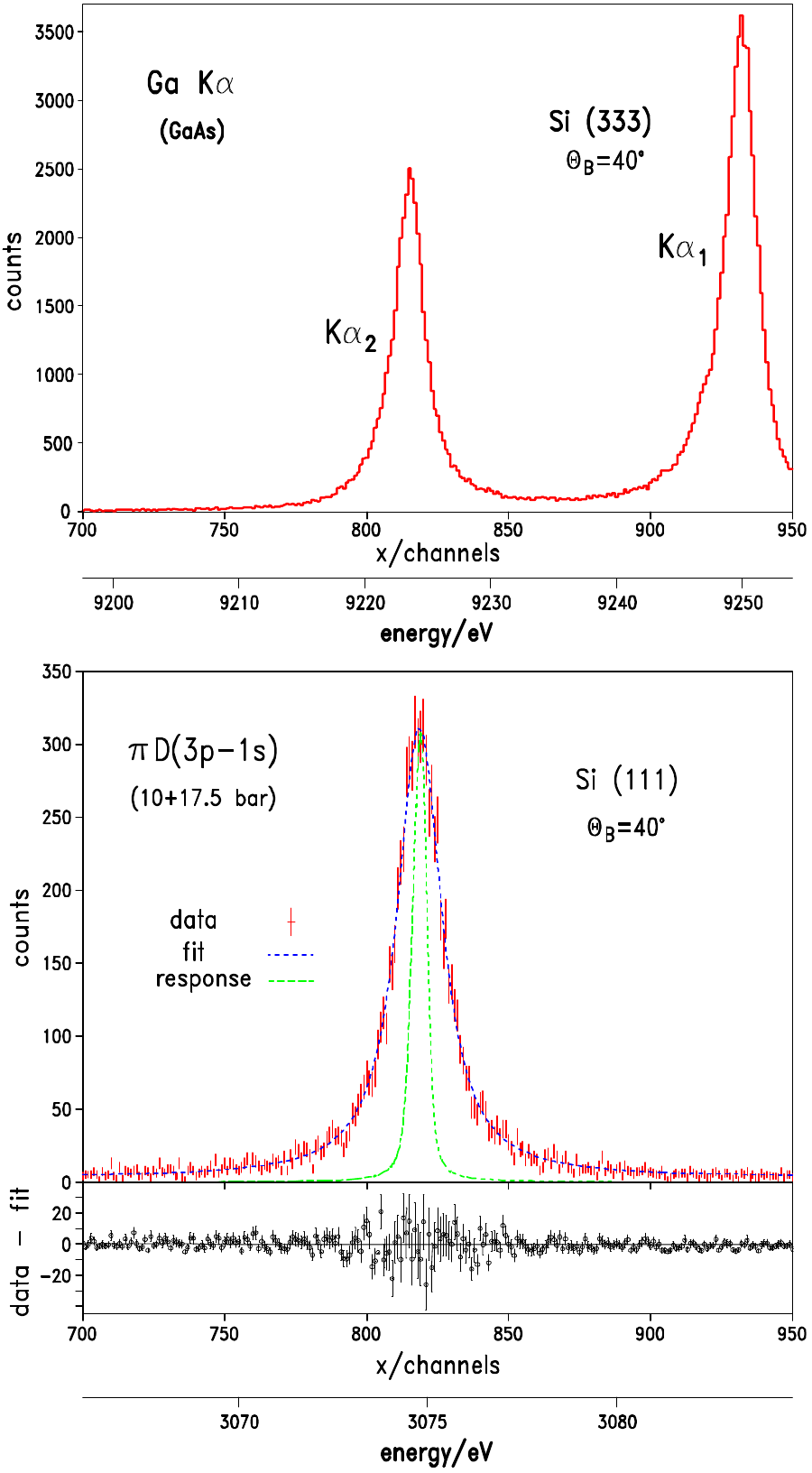}}
\caption{Top -- Ga K$\alpha_2$ calibration line. The K$\alpha_1$ component is slightly reduced by the asymmetric 
         emission characteristics of the X-ray tube.
         Middle -- sum spectrum of the $\pi$D$(3p-1s)$ transition measured at the equivalent densities of 10 and 17.5\,bar. 
         The narrow structure represents the response function of the crystal spectrometer (normalised to the height of the 
         measured spectrum).
         Bottom -- residuum of data and fit. The fit is performed without assuming any Doppler components.}
\label{figure:GaKa_piD10bar}
\end{center}
\end{figure}

\item {\bf Penetration depth.} 
The difference in penetration depth into the crystal leads to an offset of higher energetic X-rays to smaller 
diffraction angles, which has been corrected according to the approach of ref.\,\cite{Cem92}. The penetration depths 
(including absorption) were obtained from the code XOP to $0.72\,\mu$m and $3.84\,\mu$m for 3.075 and 9.225\,keV, 
respectively.
\item {\bf Fluorescence-target illumination.} 
The intensity \linebreak
profile of the X-ray tube, used to illuminate the GaAs target, was found to decrease in the direction 
of dispersion towards larger Bragg angles. Therefore, the centre of gravity of the line is shifted towards 
smaller Bragg angles or higher energies. The shape of the intensity distribution has been measured precisely and 
used for a Monte Carlo simulation to quantify the shift.
\item {\bf Mechanical stability.} 
Due to a fixed sequence for refilling the X-ray detector with liquid nitrogen (twice per day) and the gallium 
calibration (once per day after filling), small mechanical distortions arose from the cold nitrogen gas. These 
distortions were monitored every 15\,min with two inclination sensors (of 1\,$\mu$rad precison), one attached 
to the crystal chamber and the second to the detector cryostat. In this way, both short and long term movements 
of the vertical crystal axis relative to the X-ray detector were recognized. For each of the three measurement 
periods an average correction was determined from the inclinometer data. The maximal deviations were found to 
be about $\pm$\,0.1\,mrad corresponding to about one pixel.
\item {\bf Curvature correction.} 
The curvature of the reflection is determined from a parabola fit both to the $\pi$D and the Ga K$\alpha_2$ 
line, which can be assumed to be equal within the envisaged accuracy. The centres of gravity are determined 
choosing a left and right boundary aligned with the curvature of the reflection itself, the reflection divided 
vertically in 12 slices per CCD, and iterated until the minimal $\chi^2$ is reached. The maximum position by the 
various choices for the line and the boundaries are summarised in the given uncertainty.
\item {\bf Defocussing.} 
The small difference in the focal length of the $\pi$D and the Ga K$\alpha_2$ leads to a small defocussing 
correction, which was determined by means of a Monte-Carlo simulation.
\item {\bf Focal length.} 
The parts constituting the mechanical connection between crystal and detector were adjusted to the assumed 
focal length of the  $\pi$D$(3p-1s)$ transition. Their length and the non parallelity of the flanges were 
precisely surveyed resulting in an uncertainty of the distance of $\pm\,0.5$\,mm.
\item {\bf Alignment.} 
A possible deviation from the detector surface from being perpendicular to the direction crystal centre to 
detector centre leads to a correction for the measured distance of the $\pi$D and Ga K$\alpha_2$ reflection. 
The uncertainty is composed of the setup of the CCD array inside the cryostat and the deviation from being 
parallel of the flanges of the connecting tubes, which were obtained from a survey measurement.
\item {\bf Height alignment.} 
If X-ray source, bragg crystal and detector centre are aligned perfectly in-plane, a symmetric reflection with 
respect to this plane occurs at the detector surface. A misalignment causes a tilt of the reflection 
inducing a small error using the parabola approach for the curvature correction. The vertical positions of 
target, crystal, and detector centre relative to the incoming pion beam were measured to be 206.5, 205.1, and 
206.1\,mm with an accuracy of $\pm$\,0.2\,mm.
\item {\bf Pixel size.} 
The average pixel size of the CCDs was determined in a separate experiment to be $(39.9775\,\pm\,0.0006)\,\mu$m 
at the operating temperature of $-100^{\circ}$C \cite{Ind06}.
\item {\bf Temperature normalisation.} 
The crystal temperature was monitored regularly during the measurements. Though of minor impact, the lattice 
constant tabulated for $22^{\circ}$C, was readjusted to the average temperature of $30^{\circ}$C. The temperature 
was found to be stable within less than $\pm\,1^{\circ}$C. Hence, any position variation due to the change 
of the lattice constant can be neglected.
\end{itemize}
As described above, the total uncertainty in the determination of the $\pi$D$(3p-1s)$ transition energy is dominated 
by the one of the Ga calibration line ($\pm$\,27\,meV). The systematic error of about 6\,meV is about 50\% of 
the statistical uncertainty (sect.\,\ref{subsec:energy}) and comparable to the uncertainty of the calculated 
electromagnetic transition energy (sect.\,\ref{subsec:QEDenergy}).

\setlength{\tabcolsep}{1mm}
\begin{table}[t]
\begin{center}
 \caption[Corrections and errors]
   {Angular corrections (in seconds of arc (")) with associated uncertainties as well as other sources of systematic and calibration 
    errors occuring in the determination of the $\pi$D$(3p-1s)$ transition energy and hadronic shift $\epsilon_{1s}$. The total angular 
    correction being the sum of the individual terms and associated uncertainties are converted to meV. The total systematic error owing to 
    experiment and setup (errors are assumed to be uncorrelated) is given conservatively as the quadratic sum and is calculated assuming on average 
    the mechanical stability of the 17.5\,bar measurement. For comparison, the statistical error and the uncertainty of the QED transition 
    energy (see sects.\,\ref{subsec:energy} and \ref{subsec:QEDenergy}) is shown together with the one of the Ga calibration line.}
 \label{table:en_corr}
\begin{tabular}{lccccc}
  \hline\\[-3mm]
   uncertainty\          & $\pi $D  &Ga K$\alpha$ &\multicolumn{3}{c}{total correction}\\
   correction            & /\,"     & /\,"        &\multicolumn{3}{c}{/\,meV}          \\
 \hline\\[-2mm]
  index of refraction    &---       &2.4          &-42.6      &$\pm$&\,0.4\\[1mm]
  index of refraction    &21.8      &---          &387.2      &$\pm$&\,3.9\\[1mm]
  crystal bending        &-3.73     &-3.65        &-1.4       &${+\atop -}$&${0.4\atop 0.9}$\\[2mm]
  penetration depth      &-0.06     &-0.32        &4.6        &${+\atop -}$&${1.1\atop 2.7}$\\[2mm]
  illumination GaAs target &0         &-1.075       &19.1       &$\pm$&\,3.8\\[1mm]
  mechanical stability 3.3\,bar  &0.24  &         &4.2        &$\pm$&\,2.5\\[0mm]
  mechanical stability 10\,bar   &0.71  &         &12.6       &$\pm$&\,4.0\\[0mm]
  mechanical stability 17.5\,bar &0.80  &         &14.2       &$\pm$&\,1.3\\[1mm]
  curvature correction   &          &             &           &$\pm$&\,2.0\\[1mm]
  defocussing            &0         &-0.006       &0.11       &$\pm$&\,0.01\\[1mm]
  focal length           &          &              &          &$\pm$&\,0.05\\[2mm]
  alignment crystal-detector &      &             &           &${+\atop -}$&${0\atop 0.015}$\\[2mm]
  height alignment       &          &             &           &${+\atop -}$&${0\atop 0.016}$\\[2mm]
  pixel size             &          &             &           &$\pm$&\,0.14\\[1mm]
  temperature renormalisation &     &             &           &$\pm$&\,0.02\\[1mm]
  bias due to asymmetric errors &     &             & -0.4      &\\[1mm]
\hline\\[-2mm]
  total systematic error &          &             &           &${+\atop -}$&${6.0\atop 6.6}$\\[1mm]
\hline\\[-2mm]
  statistical error      &          &             &           &$\pm$&\,10.9\\[1mm]
  E$_\mathrm{QED}$       &          &             &           &$\pm$&\,7.9\\[1mm]
  Ga K$\alpha_2$         &          &             &           &$\pm$&\,27\\[1mm]
  \hline
  \end{tabular}
 \end{center}
\end{table}

\subsection{Line width}
\label{subsec:line_width} 
For comparison with data, the line shape is constructed by folding the response function (see sect.\,\ref{sec:exp} 
and fig.\,\ref{figure:GaKa_piD10bar}) with the Doppler induced width derived from a given kinetic energy distribution 
({\it e.g.}, as shown in fig.\,\ref{figure:Tkin_ESCM}), and a Lorentzian representing the natural width. Based on the 
experience of the study of the $\mu$H$(3p-1s)$ line shape\,\cite{Cov09}, in a cascade model free approach, 
the kinetic energy distribution is approximated by narrow intervals of a few eV width at 
energies which are inspired by the energy release of Coulomb de-excitation transitions. Such an energy spectrum 
assuming only a low-energy (0-2\,eV) and one high-energy component (owing to the $(4-3)$ Coulomb de-excitation) 
is included in fig.\,\ref{figure:Tkin_ESCM}. 

Spectra constructed in this way are compared to data by means of a $\chi^2$ analysis using the MINUIT 
package\,\cite{Jam75}. Natural line width, total intensity of the line, background, and relative weight of 
the Doppler contributions are free parameters of the fit. The sum of the Doppler contributions is normalised to one.

\begin{figure}[t]
\begin{center}
\resizebox{0.48\textwidth}{!}{\includegraphics{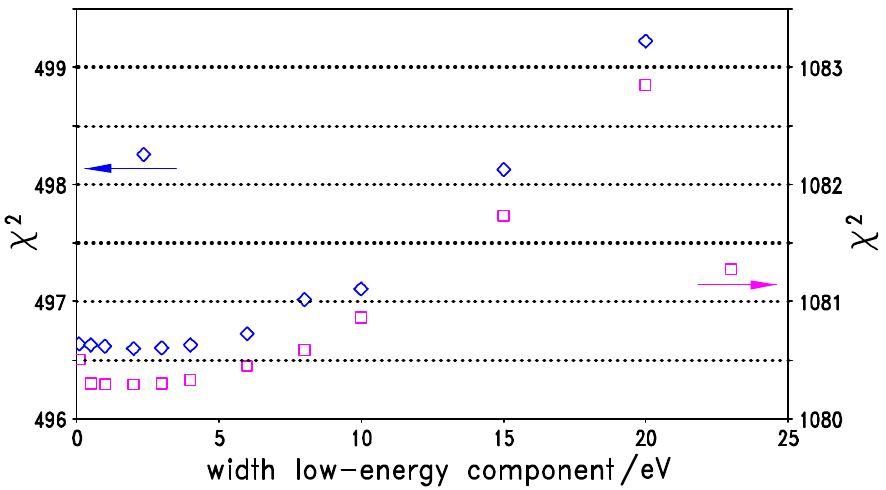}}
\caption{Search for evidence of an extended low-energy component in the kinetic energy distribution by including in the fit of the  
         data a uniform kinetic energy distribution of variable width starting at T$_{\mathrm{kin}}=0$. 
         Allowing for kinetic energies above about 8\,eV increasingly downgrades the quality of the fit.
         Diamonds (left $\chi^2$ scale) are due to the 10\,bar data. Squares (right scale) are from a 
         combined fit to the two spectra taken at 10 and 17.5\,bar. Degrees of freedom in the fits 
         are 487 (left scale) and 974 (right).}
\label{figure:searchbox1}
\end{center}
\end{figure}

Following again the approach from the analysis of the $\mu$H$(3p-1s)$ transition, one tries to identify consecutively 
individual Doppler contributions by using first a single kinetic-energy component starting at energy zero and variable width. 
This corresponds to $\pi$D systems not being accelerated or already moderated down again by collisions to energies of 
a few eV. The $\chi^2$ analysis shows that the upper boundary of the low-energy contribution must not exceed 
8\,eV. This result was achieved independently for the spectra taken at 10\,bar and the 17.5\,bar equivalent density 
(Fig.\,\ref{figure:searchbox1}). The result for the natural line width $\Gamma_{1s}$ turned out to be insensitive to 
the upper boundary when keeping its value at 8\,eV or below. Therefore, the low-energy component was fixed to the 
interval 0\,--\,2\,eV in the further analysis. 

The most important higher energetic components are expected at about 80 and 115\,eV stemming from the\linebreak 
Coulomb de-excitation transitions $(4-3)$ and $(5-3)$. Searches for any of these contributions failed, 
even when using the sum spectrum of the two measurements at 10 and 17.5\,bar. Varying energy or width 
of the high-energy Doppler contribution did not change this result.

In this respect, pionic deuterium differs significantly from pionic and muonic hydrogen. There, such high-energy 
contributions were clearly identified from the width analysis of the X-ray transitions\,\cite{Sch01,Got08,Cov09}. 
For that reason, the ESCM calculation of the kinetic energy distribution for the $\pi$H$(3p-1s)$ case, 
scaled to $\pi$D kinematics (Fig.\,\ref{figure:Tkin_ESCM}), is unable to reproduce the $\pi$D$(3p-1s)$ 
line shape, because it contains rather strong contributions from $(4-3)$ and $(5-3)$ Coulomb de-excitation. 
At present, no explanation has been found for such an unequal behaviour.

\begin{figure}[b]
\begin{center}
\resizebox{0.45\textwidth}{!}{\includegraphics{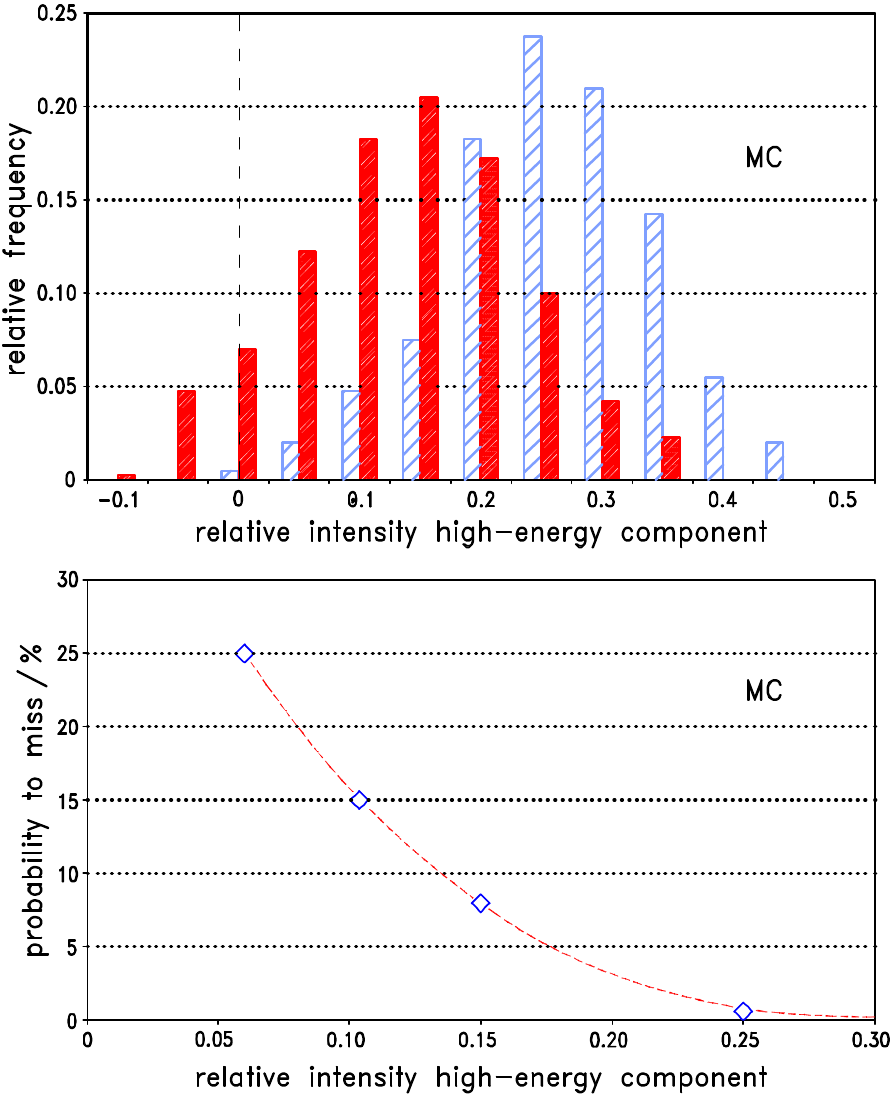}}
\caption{Top -- distributions of relative intensity of a Doppler broadened component induced by the $(4-3)$ Coulomb 
         de-excitation transition found from fits to Monte-Carlo (MC) generated spectra. The relative strength used 
         for the simulations is 10\% (filled) and 25\% (hatched) relative intensity, respectively. In both cases, 
         the result is based on 400 spectra generated for the statistics of the sum of the 10 and 17.5\,bar data. 
         Bottom -- probability to miss a Doppler contribution corresponding to the $(4-3)$ Coulomb de-excitation 
         (modeled by a uniform kinetic energy distribution from ($76-84)$\,eV) as a function of its relative 
         intensity.}
\label{figure:sensitivity}
\end{center}
\end{figure}

It has been studied in detail which fraction of high-energy components may be missed in the fit by studying 
Monte-Carlo generated spectra for the statistics collected for the sum of the 10 and 17.5\,bar measurements. 
The intensity of the Doppler contributions was not restricted to positive values to allow an unbiased search 
for the minimum $\chi^2$ (Fig.\,\ref{figure:sensitivity} -- top). The normalization to one is then maintained 
by a correspondingly increased value for the other component.

The probability to miss a Doppler component, stemming from kinetic energies around 80\,eV and corresponding to 
the $(4-3)$ transition, is displayed in fig.\,\ref{figure:sensitivity} -- bottom as a function of its relative 
strength. It can be seen, that a contribution of 25\% or larger can hardly be missed. For 10\% relative intensity, 
the chance is about 15\% to find intensities $\leq 0$ in the fit. Taking symmetric limits around the maximum at 
about 10\% relative frequency (Fig.\,\ref{figure:sensitivity} -- top) corresponds to 1$\sigma$ with respect to 
the full distribution. For each set of conditions 400 simulations were performed. 

\begin{figure}[h]
\begin{center}
\resizebox{0.48\textwidth}{!}{\includegraphics{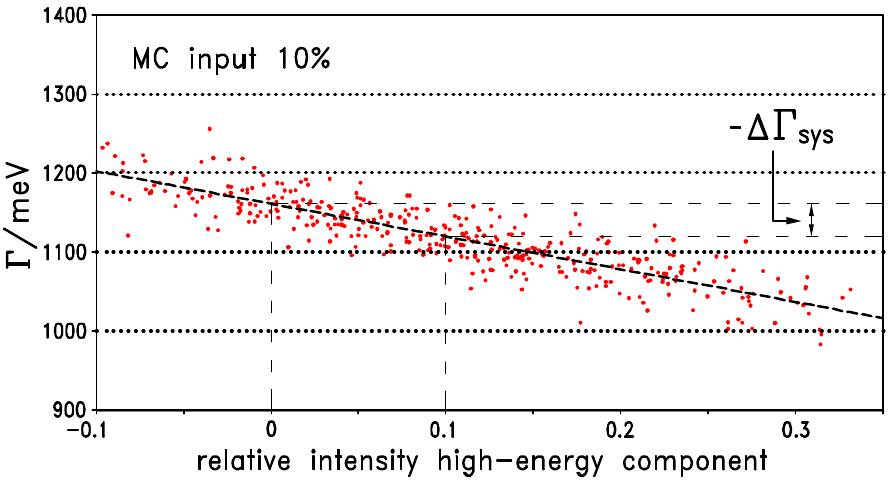}}
\caption{Distributions of the Lorentz width $\Gamma$ {\it versus} the relative intensity of the Doppler broadened component      
         stemming from the $(4-3)$ Coulomb de-excitation. The results are obtained from 400 Monte-Carlo simulations 
         performed for the statistics of the sum of the 10 and 17.5\,bar data. Input values are 10\% relative intensity 
         for the Doppler affected component and 1130\,eV for $\Gamma$.}
\label{figure:wolke}
\end{center}
\end{figure}

Assuming no Doppler components yields an upper limit for $\Gamma_{1s}$ which is identical to the result obtained when 
using only the low-energy component of 0--2\,eV. The residuum of such a fit is shown in fig.\,\ref{figure:GaKa_piD10bar} -- bottom. 
The limit of sensitivity for a component at 80\,eV kinetic energy of 
10\% yields a lower bound for $\Gamma_{1s}$ ($-\bigtriangleup\Gamma_{sys}$) corresponding to the above-mentioned 
1$\sigma$ criterium (Fig.\,\ref{figure:sensitivity}). The distribution of the results for the weight of this component 
and the Lorentz width $\Gamma$ reflects the fluctuations owing to the limited statistics (Fig.\,\ref{figure:wolke}). 

From the various above-mentioned sets, each with 400 simulations, a possible systematic deviation (bias) of the 
analysis code was examined. Such a deviation stems from an imperfect description of the probability distribution 
by the fit model\,\cite{Ber02}, here used to extract $\Gamma_{1s}$, which becomes more and more important with 
decreasing statistics. Where for the individual spectra measured at 3.3, 10, and 17.5\,bar the bias was found to 
depend significantly on the statistics ((-38$\pm$3), (17$\pm$2), and (11$\pm$2)\,meV), for the sum spectrum 
(10+17.5)\,bar it becomes  almost negligible ((-2$\pm$2)\,meV). The error of $\approx 2$\,meV of the bias is given 
by the average of the results from the 400 Monte-Carlo spectra, which in turn determines the number of simulations 
necessary. 

The behaviour of the bias is understood as follows. For lower statistics, 
the suppression of the tails results in smaller widths, whereas in the case of higher statistics the fit tends to 
include background into the tail. The background has been found to be constant in all spectra. The systematical 
uncertainty of $\Gamma_{1s}$ due to different background levels within the limits as obtained from the fit to 
the data is estimated to below 1\,meV by means of Monte Carlo simulations. 

Again 400 Monte-Carlo generated spectra, based on a calculated ESCM energy distribution, were used to 
quantify the sensitivity to any further high-energy component. An additional 6\% fraction corresponding to 
the predicted strength of the $(5-3)$ de-excitation together with a 10\% contribution from the $(4-3)$ is 
identified by the fit in more than 2/3 of all cases. In the other 1/3 of the fits, the $(5-3)$ fraction is 
absorbed into the $(4-3)$ contribution leading to an overestimate of its intensity and a shift to higher energies. 

No evidence for an additional, {\it i.e.}, asymmetric broadening from non resolved satellites due to molecular formation has been found within 
the experimental accuracy. This is corroborated by the observation that a possible density dependence of the $\pi$D$(3p-1s)$ 
energy contradicts the expected behaviour (see sects.\,\ref{subsec:energy} and \ref{subsec:shift}).

\section{Results and discussion}\label{sec:results}

\subsection{Experimental $\pi$D$(3p-1s)$ transition energy}\label{subsec:energy} 

The results for the $\pi$D$(3p-1s)$ X-ray energy are summarized in table\,\ref{table:energy} taking already 
into account the corrections and their uncertainties as given in table\,\ref{table:en_corr}. The values 
obtained for the three different target densities are consistent within two standard deviations (neglecting 
the common uncertainty of the Ga K$\alpha_2$ energy).

It is concluded that we cannot identify radiative de-excitation from molecular states within the experimental 
accuracy, even more, as it is expected that such a process decreases the $\pi$D transition energy with density.
Noteworthy, that also the analysis of the $\mu$H$(3p-1s)$ line shape yields no evidence for a broadening 
owing to molecular effects\,\cite{Cov09}.

Inspecting the background at the low-energy side of the $\pi$D transition for possible Auger stabilised molecules, 
a weak structure was found separated by $(28.6\,\pm\,0.5)$\,eV. Its relative intensity was determined to 
be $(3.4\,\pm\,1.6)\cdot 10^{-3}$ of the main line.

A weighted average is calculated for the $\pi$D$(3p-1s)$ transition energy from the results for the three 
different target densities by using the individual contributions to the statistical and systematical error. 
The final value and error (quadratic sum) is obtained when combining these errors with the uncertainty of 
$\mathrm{\Delta}$E$_{\mathrm{Ga}}$ for the Ga\,K$\alpha_2$ line: 
\begin{eqnarray}
\mathrm{E}_{\pi\mathrm{D}(3p-1s)}=(3075.583\,\pm\,0.030)~\mathrm{eV}. 
\end{eqnarray}
The experimental uncertainty is dominated by the error of the Ga K$\alpha_2$ calibration line. In case a 
significant theoretical progress requires further improvement, the calibration error could be reduced by 
a factor of about 2 in a dedicated measurement using a double flat crystal spectrometer.

\setlength{\tabcolsep}{1.75mm}
\begin{table}[h]
 \begin{center}
 \caption{Measured $\pi$D$(3p-1s)$ transition energy and associated uncertainties.}
  \label{table:energy}
 \begin{tabular}{ccccc}
\hline\\[-3mm]
 equivalent         & E$(3p-1s)$ & \multicolumn{3}{c}{$\mathrm{\Delta}\mathrm{E}(3p-1s)$}\\
 density            &            & stat           & sys              & $\Delta$E$_{\mathrm{Ga}}$  \\[0mm]
 /bar               & /\,eV      & /\,meV         & /\,meV           & /\,meV \\[0mm]
 \hline\\[-2mm]
  3.3               &3075.509    &$\pm $28        &${+\,6\atop -7}$  &$\pm $27\\[1mm]
  10                &3075.594    &$\pm $17        &${+\,7\atop -8}$  &$\pm $27\\[1mm]
  17.5              &3075.599    &$\pm $16        &${+\,6\atop -7}$  &$\pm $27\\[1mm]
   \hline\\[-3mm]
  weighted average  &3075.583    &$\pm $11        &$\pm $7           &$\pm $27\\[0mm]
   \hline
  \end{tabular}
 \end{center}
\end{table}

\subsection{Electromagnetic $\pi$D$(3p-1s)$ transition energy}\label{subsec:QEDenergy} 

In order to extract the hadronic shift, the pure electromagnetic transition energy has been recalculated within the 
framework of quantum electrodynamics (QED). The contributions to the level energies are given in table \ref{table:qed}. 
The error of the calculated QED transition energy E$_\mathrm{QED}$ = 3077.939$\,\pm\,$0.008\,eV is dominated by the 
uncertainty of the charged pion mass\,\cite{PDG08} ($\pm$\,7.72\,meV), whereas the ones of the charge radii of the 
deuteron ($\pm$\,1.19\,meV) and the pion\,\cite{PDG08} ($\pm$\,0.95\,meV) contribute only marginal. The numerical 
accuracy is better than 1\,meV. For this calculation, a deuteron charge radius $r_{\mathrm{D}}=(2.12809\,\pm\,0.00031)$\,fm 
has been used as derived from the most recent value of the proton charge radius\,\cite{Poh10} and the radii 
difference obtained from the hydrogen-deuterium isotope shift measured by means of two-photon 
spectroscopy\,\cite{Par10}.

The correction to the ground-state energy arising from the electric polarizability of the deuteron 
of (-26.3$\pm$0.5) meV is obtained by scaling from a calculation for muonic deuterium\,\cite{Ros93,Ros95}. 
Contributions from the polarizability of the pion\,\cite{Ahr05} are three orders of magnitude smaller 
and, consequently, are neglected.

\setlength{\tabcolsep}{0.4mm}
\begin{table}[t]
 \begin{center}
 \caption[Contributions to $E_{QED}$]
   {Contributions to the pure electromagnetic binding and the $(3p-1s)$ transition energies. The terms 
    vac. pol. 11, 13, and 21 stand for Uehling, Wichmann-Kroll, and K\"all\'en-Sabry contributions. The shift value 
    due to the polarizability of the deuteron is derived from results given in refs.\,\cite{Ros93,Ros95}. A "0" 
    indicates a negligibly small but finite value.}
 \label{table:qed}
 \begin{tabular}{lccc}
 \hline\\[-3mm]
                          &$3p$      &$1s$        & E$_\mathrm{QED}(3p-1s)$\\[1mm]
                          & /\,eV    & /\,eV      & \,/eV\\[0mm]
 \hline\\[-3mm]
 Coulomb                  &-384.31079& -3458.52422& 3074.21344\\
 self-energy\,+\,finite size &     0 &    ~0.00220&   -0.00220\\
 vac. pol. 11             & -0.01432 &    -3.72948&   ~3.71516\\
 muon vac. pol. 11        &        0 &    -0.00034&   -0.00034\\
 vac. pol. 13             &        0 &    ~0.00002&   -0.00002\\
 vac. pol. 21             & -0.00013 &    -0.02793&   ~0.02780\\
 recoil 1                 & -0.00004 &    -0.00297&   ~0.00293\\
 relativistic recoil      &        0 &    ~0.00029&   -0.00029\\
 two-loop vac. pol.       &        0 &    -0.00577&   ~0.00577\\
 pion finite size         &        0 &    -0.04764&   ~0.04764\\
 deuteron polarizability  &     0    &    -0.0263\,~  &  ~0.0263\,~~\\
 $\epsilon_{3p}$          & -0.000008&   ---      &  ~~~~0.000008\,~~\\
 atom recoil              &    ---   &   ---      &   -0.00235\\
\hline\\[-3mm]
 total                    &-384.32528& -3462.26684& 3077.93922\\
  \hline
  \end{tabular}
 \end{center}
\end{table}   

The strong-interaction shift of the $\pi$D $3p$ state is estimated as for the s-wave in leading 
order by the sum of the pion-proton and pion-neutron interaction. The $\pi d$ scattering volume may be 
expressed by twice the isoscalar and angular momentum averaged parameter $c_0$ of the Kisslinger 
potential\,\cite{Eri88-6,Bar97,Hue75}, which can be calculated from the $\pi N$ scattering volumes. Values 
are given, {\it e.g.}, in ref.\,\cite{Koc86}. Inserting the result for $2\cdot c_0$ in the Trueman formula 
for the $3p$ state\,\cite{Tru61,Lam6970}, one obtains a negligibly small shift of $\epsilon_{3p}\approx 8\,\mu$eV.

\subsection{Strong-interaction shift $\epsilon_{1s}$}\label{subsec:shift} 

Subtracting measured and calculated QED transition energies, the hadronic shift is obtained to 
\begin{eqnarray}
\epsilon_{1s}=\,-\,(2356\,\pm\,31)\,\mathrm{meV}\label{eq:eps_final}. 
\end{eqnarray}
The negative sign indicates a repulsive interaction, {\it i.e.}, the strong interaction reduces the atomic 1s level 
binding energy. 

\begin{figure}[t]
\begin{center}
\resizebox{0.48\textwidth}{!}{\includegraphics{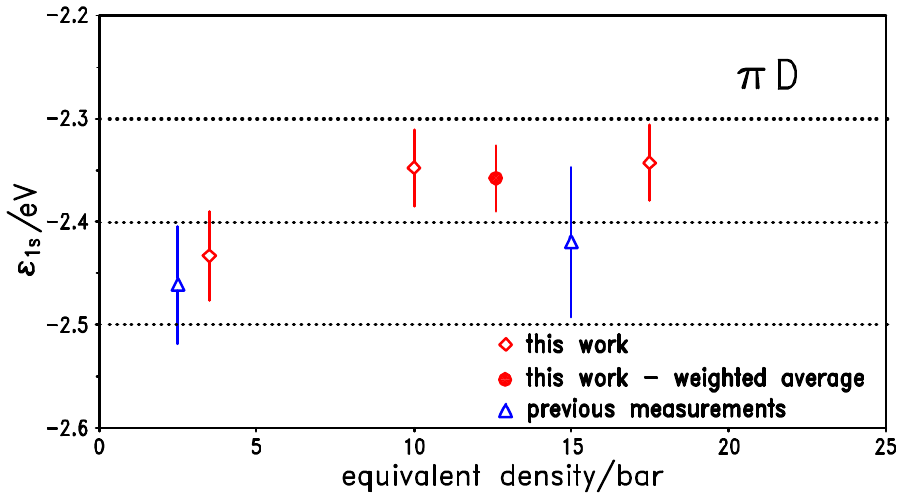}}
\caption{Experimental information on the hadronic shift in pionic deuterium. The error of the individual measurements at 
         3.3, 10, and 17.5\,bar equivalent density and, therefore, also of the weighted average is dominated by the uncertainty of the 
         Ga K$\alpha_2$ energy. The results from the two previous experiments at 2.5 bar \cite{Hau98} and 15\,bar  \cite{Cha9597} 
         are adjusted according to the new values for the pure electromagnetic transition energies (Table\,\ref{table:piD_results}).}
\label{figure:shift}
\end{center}
\end{figure}

Our result is compared to previous measurements in table\,\ref{table:piD_results} and 
fig.\,\ref{figure:shift}. Note that the electromagnetic transition energy E$_\mathrm{QED}$ differs slightly in 
the analysis of the various experiments. Worth mentioning, that the results of the three precision experiments 
are in good agreement though three different calibration lines have been used.

Combining the results of this experiment and the ones given in refs. \cite{Cha9597} and \cite{Hau98} yields a weighted average of 
$\epsilon_{1s}=(-2379 \pm\,19)$\,meV assuming any absence of a density dependence for the $\pi$D$(3p-1s)$ transition energy. 

Inspecting fig.\,\ref{figure:shift} suggests that an extrapolation to density zero yields the value aimed at. A linear 
fit yields a slope of $(6.3\pm 1.7)$\,meV/bar and $\epsilon _{1s}=(2445\pm 27)$\,meV at density zero. However, at present 
there is no explanation for such a density dependence (see sect.\,\ref{sec:cascade}), which might be subject for a next 
generation experiment and, therefore, we stick to the density independent average given in (\ref{eq:eps_final}).

\subsection{Strong-interaction width $\Gamma_{1s}$}\label{subsec:width} 

Corresponding to the discussion in sect.\,\ref{subsec:line_width}, the hadronic broadening is extracted by introducing 
only a low-energy component from $0-2$\,eV in the kinetic energy distribution. The combined result is obtained 
by averaging according to the statistical weight (Table\,\ref{table:width}). The large negative error is dominated 
by the uncertainty that a Doppler contribution of the level of about 10\% might be missed in the analysis.  

\setlength{\tabcolsep}{4.5mm}
\begin{table}[b]
 \begin{center}
 \caption{Pionic deuterium ground state width $\Gamma_{1s}$ as obtained from the individual 
          measurements at 3.3, 10, and 17.5\,bar and the sum spectrum at (10+17.5)\,bar. 
          The weighted average is calculated from the result of the sum spectrum (10+17.5)\,bar 
          and the one of the 3.3\,bar measurement.}
  \label{table:width}
\begin{tabular}{cccc}
 \hline\\[-3mm]
 equivalent        & $\Gamma_{1s}$ & \multicolumn{2}{c}{$\Delta\Gamma_{1s}$} \\
 density           &               &          stat              &      sys   \\
  /\,bar           & /\,meV        & /\,meV                     & /\,meV \\
 \hline\\[-2mm]
   3.3             & 1246          & $\pm$\,71       &${+\,~0~~\atop-\,~152}$ \\[2mm]
   10              & 1177          & $\pm$\,38       &${+\,~0~~\atop-\,~92}$ \\[2mm]
   17.5            & 1121          & $\pm$\,32       &${+\,~0~~\atop-\,~81}$ \\[1mm]
   \hline\\[-2mm]
   $(10+17.5)$     & 1162          & $\pm$\,24       &${+\,~0~~\atop-\,~43}$ \\[1mm]
   \hline\\[-2mm]
 weighted average  & 1171          & $\pm$\,23       &${+\,~0~~\atop-\,~43}$ \\[1mm]
  \hline
   \end{tabular}
 \end{center}
\end{table}   
\setlength{\tabcolsep}{1.2mm}
\begin{table*}[t]
\begin{center}
\caption{Transition energies and hadronic effects found for pionic deuterium and corresponding experimental conditions. 
         The total error of the measured transition energy E$_{\mathrm{exp}}$ is calculated quadratically 
         from the statistical and systematical contributions. The hadronic shift is defined by 
         $\epsilon_{1s}\equiv \mathrm{E_{exp}-E_{QED}}$.
         Note the change of E$_\mathrm{QED}$ ($^{*}$) due to a new calculation performed for the analysis of this experiment   
         which shifts the result of \cite{Cha9597} to $\epsilon_{1s}=-2419\pm 100$\,meV. Also for  the $(2p-1s)$ transition a new 
         QED value of 2597.519$\pm$0.008\,eV is obtained resulting in $\epsilon_{1s}=-2461\pm 55$\,meV for the experiment 
         of\,\cite{Hau98}. No uncertainties for the QED values have been reported for the experiments decribed in 
         refs.\,\cite{Hau98,Bov81,Bov85}.}
\label{table:piD_results}
\begin{tabular}{ccclllrclrclrclc}
\hline\\[-2mm]

transition     & equivalent  &energy        &\multicolumn{3}{c}{E$_{\mathrm{exp}}$}&\multicolumn{3}{c}{E$_\mathrm{QED}$}&\multicolumn{3}{c}{$\epsilon_{1s}$}   &\multicolumn{3}{c}{$\Gamma_{1s}$}& \\ 
               & density     &calibration   &              &              &                                    &                                      &                                 & \\
               &  /\,bar     &              &\multicolumn{3}{c}{/\,eV}    &\multicolumn{3}{c}{/\,eV}           &\multicolumn{3}{c}{/\,meV}            &\multicolumn{3}{c}{/\,meV}       & \\
\hline\\[-2mm]
$\pi D(2p-1s)$&$4-8.5$      &Bi M$_V$\,edge&2592.8  &${+\atop-}$&${1.6\atop2.0}$&2597.61&$\pm$&0.15           &--\,4800&${+\atop-}$&${1600\atop2000}$&     &---&                       &\cite{Bai74}\\[1mm]
$\pi D(2p-1s)$&$\approx 3.5$&Cu K$\alpha$  &2593.3  &$\pm$&2.3           &\multicolumn{3}{c}{2598.1~~}        &--\,4800&$\pm$&2300                   &     &---&                       &\cite{Bov81}\\
$\pi D(2p-1s)$&$\approx 50$ &Cl K$\alpha$  &2592.1  &$\pm$&0.9           &\multicolumn{3}{c}{2597.6~~}        &--\,5500&$\pm$&900                    &     &---&                       &\cite{Bov85}\\
$\pi D(3p-1s)$&15           &Ar K$\alpha$  &3075.52 &$\pm$&0.07                 &3077.95&$\pm$&0.01$^{*}$     &--\,2430&$\pm$&100                    & 1020&$\pm$&210                  &\cite{Cha9597}\\
$\pi D(2p-1s)$&2.5          &Cl K$\alpha$  &2595.058&$\pm$&0.055         &\multicolumn{3}{c}{~~2597.527$^{*}$}&--\,2469&$\pm$&55                     & 1194&$\pm$&105                  &\cite{Hau98}\\[0.5mm]
$\pi D(3p-1s)$&3.3/10/17.5  &Ga K$\alpha_2$&3075.583&$\pm$&0.030         &3077.939&$\pm$&0.008$^{*}$          &--\,2356&$\pm$&31                     & 1171&${+\atop-}$&${23\atop49}$  &{\em this exp.}\\[1mm]
\hline\\[-4mm]
\end{tabular}
\end{center}
\end{table*}

Combining all errors (quadratic sum), we obtain
\begin{eqnarray}
\Gamma_{1s}=\left(1171\,{+\,~23\atop-\,~49}\right)\,\,\mathrm{meV}\label{eq:Ga_final}. 
\end{eqnarray}

\begin{figure}[h]
\begin{center}
\resizebox{0.48\textwidth}{!}{\includegraphics{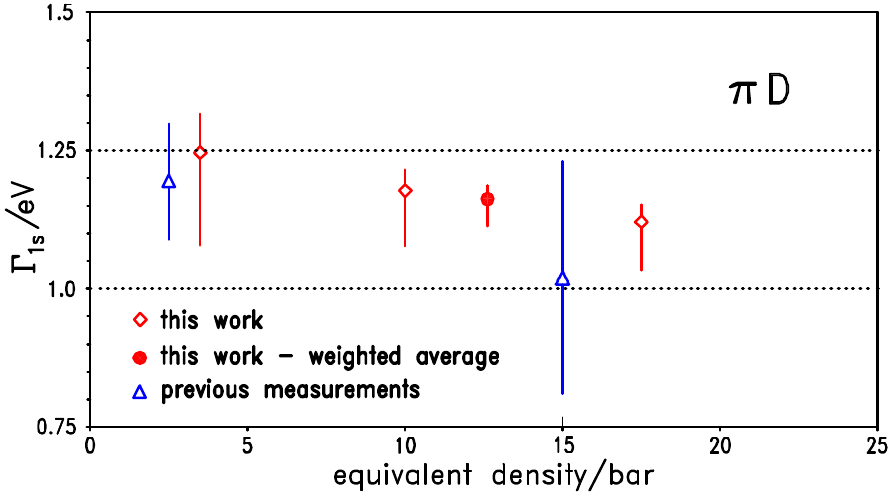}}
\caption{Experimental information on the hadronic broadening in pionic deuterium. 
         Previous measurements were performed at 2.5 bar \cite{Hau98} and 15\,bar \cite{Cha9597} equivalent density.}
\label{figure:width}
\end{center}
\end{figure}

The result is in good agreement with the earlier measurements, but a factor of about 3 more precise 
(Table\,\ref{table:piD_results} and fig.\,\ref{figure:width}). The weighted average of the results of this 
experiment, \cite{Cha9597}, and \cite{Hau98} is $\Gamma_{1s}=\left(1165\,{+\,~22\atop-\,~38}\right)$\,meV. 

Again, a possible density dependence might be suspected (Fig.\ref{figure:width}). Performing a linear fit---taking 
into account the asymmetry of the errors---yields $\Gamma_{1s}=(1229\pm75$)\,meV at density zero and a slope of 
$(-6.8\pm 4.8)$\,meV/bar. Also here, no reason was found for such a density dependence and we keep the density 
independent result.

\subsection{Scattering length $a_{\pi\mathrm{D}}$}\label{subsec:scatteringlength} 

Evaluating the DGBT formula (\ref{eq:deser}), one obtains for the complex scattering length 
$a^{\mathrm{LO}}_{\pi\mathrm{D}}=[(-25.07\pm0.33)\,+i\,(6.23\pm0.26)]\cdot 10^{-3}\,m^{-1}_{\pi}$. Inserting these 
leading order values in eqs. (\ref{eq:true_re}) and (\ref{eq:true_im}) one obtains 
\begin{eqnarray}
\mathrm{Re}\, a_{\pi\mathrm{D}} &=& 0.010642\,m^{-1}_{\pi}/eV\cdot\,0.9968\,\cdot\,\epsilon_{1s}\label{eq:Re_corr}\\ 
\mathrm{Im}\, a_{\pi\mathrm{D}} &=& 0.010642\,m^{-1}_{\pi}/eV\cdot\,0.9989\,\cdot\,\left(\Gamma_{1s}/2\right)\,\label{eq:Im_corr}
\end{eqnarray}
with the factors 0.9968 and 0.9989 representing the "Trueman" correction. Consequently, our result 
for the $\pi$D scattering length reads
\begin{eqnarray}
a_{\pi\mathrm{D}}=\left[(-24.99\pm 0.33)+i\left(6.22{+\,0.12\atop -0.26}\right)\right]\times 10^{-3}\,m^{-1}_{\pi}.\nonumber
          \\\label{eq:apiD}
\end{eqnarray}
Real and imaginary part of the $\pi$D scattering length $a_{\pi\mathrm{D}}$ as determined from the various 
measurements and including the Trueman correction are compared in table\,\ref{table:scat}. 

The $\pi$D scattering length was studied recently within the $\chi$PT approach\,\cite{Bar11}. In this work, almost 
prefect overlap is achieved of all constraints imposed on the $\pi N$ iso\-scalar and isovector scattering length 
$a^+$ and $a^-$ by pionic hydrogen and deuterium data. In addition, in a combined analysis using the preliminary 
results for shift and width in $\pi$H as given in\,\cite{Got08} together with the new shift value in $\pi$D from this 
experiment, the uncertainty for $a^+$ could be reduced by a factor of two yielding a positive value by 
two standard deviations. By using the results for $a^+$ and $a^-$ from this combined analysis, 
$\mathrm{Re}\,a_{\pi\mathrm{D}}=(-25.4\pm 2.6)\times 10^{-3}\,m^{-1}_{\pi}$ is obtained.

The imaginary part was calculated in LO $\chi$PT to be 
$\mathrm{Im}\,a_{\pi\mathrm{D}}=(5.65\pm1.75)\cdot 10^{-3}\,m^{-1}_{\pi}$\,\cite{Len07}. One result of this work is 
that the real part induced by this absorptive contributions turns out to be $(7\pm 2)$\% of the total 
$\mathrm{Im}\, a_{\pi\mathrm{D}}$. It is much smaller than the usual naive estimate of 
$\mathrm{Re}\,a^{induced}_{\pi\mathrm{A}} \approx \mathrm{Im}\,a_{\pi\mathrm{A}}$\,\cite{Bru55}, but can be traced 
back in the case of $\pi$D to cancellations of individually sizeable terms.
 
\setlength{\tabcolsep}{1mm}
\begin{table*}[t]
\begin{center}
\caption{Real and imaginary part $\mathrm{Re}\,a_{\pi\mathrm{D}}$ and $\mathrm{Im}\,a_{\pi\mathrm{D}}$ of the $\pi$D scattering length as extracted 
         from atomic data including the Trueman correction (\ref{eq:true_re}) and (\ref{eq:true_im}).  
         Threshold parameter $\alpha$ for pion production and $\mathrm{Im}\,a_{\pi\mathrm{D}}$ are corrected for radiative capture 
         and pair conversion (\ref{eq:Ima_nr}) but omitting any isospin breaking contributions. Production experiments 
         reporting only the statistical uncertainty are marked (*). Arrows ($\leftarrow/\rightarrow $) indicate primary 
         and derived quantities. The last line shows the result of a combined analysis of $\pi$H and $\pi$D data (see 
         sect.\,\ref{subsec:scatteringlength}). Theoretical approaches are listed in chronological order. }
\label{table:scat}
\begin{tabular}{lcccccccccccc}
\hline\\[-2mm]
                      &\multicolumn{3}{c}{$\mathrm{Re}\,a_{\pi\mathrm{D}}$} &~~~~&\multicolumn{3}{c}{$\mathrm{Im}\,a_{\pi\mathrm{D}}$} &~~~~&\multicolumn{3}{c}{$\alpha$}&\\ 
                      &\multicolumn{3}{c}{$/\,10^{-3}\cdot m^{-1}_{\pi}$}&~~~~&\multicolumn{3}{c}{$/\,10^{-3}\cdot m^{-1}_{\pi}$}&~~~~&\multicolumn{3}{c}{/\,$\mu$b}&\\
\hline\\[-2mm]
\multicolumn{13}{l}{\it{Pionic deuterium}}\\[1mm]
$2p\rightarrow 1s$           &-  52   &${+\atop-}$&${22\atop17}$&&  & --- &                       &&   & --- &                    &\cite{Bai74} \\[1mm]
$2p\rightarrow 1s$           &-  52   &$\pm$&25               &&    & --- &                       &&   & --- &                    &\cite{Bov81} \\
$2p\rightarrow 1s$           &-  84   &$\pm$&9                &~~~~~&    & --- &                  &&   & --- &                    &\cite{Bov85} \\
$3p\rightarrow 1s$           &-  25.9 &$\pm$&1.1              &&5.45&$\pm$&1.12                   &$\rightarrow$&220&$\pm$&45                 &\cite{Cha9597} \\
$2p\rightarrow 1s$           &-  26.3 &$\pm$&0.6              &&6.35&$\pm$&0.56                   &$\rightarrow$&257&$\pm$&23                 &\cite{Hau98} \\[1mm]
$3p\rightarrow 1s$           &-  24.99&$\pm$&0.33             &&6.22&${+\atop-}$&${0.12\atop0.26}$&$\rightarrow$&251&${+\atop-}$&${~5\atop11}$&{\em this exp.}\\[1mm]
\hline\\[-2mm]
\multicolumn{13}{l}{\it{Pion production and absorption}}\\[1mm]
$pp\rightarrow d\pi^+$       &&&                               &&3.33&$\pm$&0.36                   &$\leftarrow$&138&$\pm$&15$^{*}$ &\cite{Cra55} \\
$pp\rightarrow d\pi^+$       &&&                               &&5.80&$\pm$&0.48                   &$\leftarrow$&240&$\pm$&20$^{*}$ &\cite{Ros67} \\
$pp\rightarrow d\pi^+$       &&&                               &&4.35&$\pm$&0.48                   &$\leftarrow$&180&$\pm$&20$^{*}$ &\cite{Ric70} \\
$pp\rightarrow d\pi^+$       &&&                               &&5.5 &$\pm$&1.1                    &$\leftarrow$&228&$\pm$&46       &\cite{Aeb76} \\
$pn\rightarrow d\pi^0$       &&&                               &&4.44&$\pm$&0.12                   &$\leftarrow$&184&$\pm$&5$^{*}$  &\cite{Hut90} \\
$d\pi^+\rightarrow pp$       &&&                               &&4.20&$\pm$&0.07                   &$\leftarrow$&174&$\pm$&3$^{*}$  &\cite{Rit91} \\
$p_{pol}p\rightarrow d\pi^+$ &&&                               &&5.02&$\pm$&0.12                   &$\leftarrow$&208&$\pm$&5$^{*}$  &\cite{Hei96} \\
$pp\rightarrow d\pi^+$       &&&                               &&4.95&$\pm$&0.22                   &$\leftarrow$&205&$\pm$&9$^{*}$  &\cite{Dro98} \\
\hline\\[-2mm]
\multicolumn{13}{l}{\it{Theoretical approach}}\\[1mm]
Watson-Brueckner             &&&                               &&&&                                &&140&$\pm$&50       &\cite{Ros54} \\
Rescattering                 &&&                               &&&&                                &&   &146&           &\cite{Kol69} \\
Rescattering                 &&&                               &&&&                                &&   &201&           &\cite{Rei69} \\
Faddeev (Reid soft core)     &&&                               &&&&                                &&   &220&           &\cite{Afn74} \\
Faddeev (Bryan-Scott)        &&&                               &&&&                                &&   &267&           &\cite{Afn74} \\
Coupled channels             &\multicolumn{3}{c}{- 30}         &&&&                                &&\multicolumn{3}{c}{90 -- 165}  &\cite{Fay80} \\
Heavy meson exchange         &&&                               &&&&                                &&203&$\pm$&21       &\cite{Hor93,Nis96} \\
Relativistic field theory    &\multicolumn{3}{c}{- 57}         &&&&                                &&\multicolumn{3}{c}{ }  &\cite{Iva97} \\
$\chi$PT NLO                 &&&                               &&5.46&$\pm$&1.74                   &$\leftarrow$&220&$\pm$&70       &\cite{Len06} \\
$\chi$PT NLO                 &&&                               &&5.65&$\pm$&1.60                   &$\rightarrow$&228&$\pm$&65      &\cite{Len07} \\
$\chi$PT analysis combining  $\pi$H and $\pi$D~~~ &- 25.4&$\pm$&2.6    &&&&    &&&&                &\cite{Bar11} \\
\hline\\[-4mm]
\end{tabular}
\end{center}
\end{table*}

\subsection{Threshold parameter $\alpha$ in pion absorption}\label{subsec:alpha} 

The value derived from the $\pi$D hadronic width $\Gamma_{1s}$ reads
\begin{eqnarray}
\alpha  &=& \left(251{+\,5\atop -11}\right)\,\mu\mathrm{b}\,.
\end{eqnarray}
The central value differs by 1\,$\mu$b from the one given in ref.\,\cite{Str10}, because there the correction 
term $\delta^{vac}_{\mathrm{D}}$ (see sect.\,\ref{sec:strong_interaction_shift}) had not been considered. 

\begin{figure}[h]
\begin{center}
\resizebox{0.48\textwidth}{!}{\includegraphics{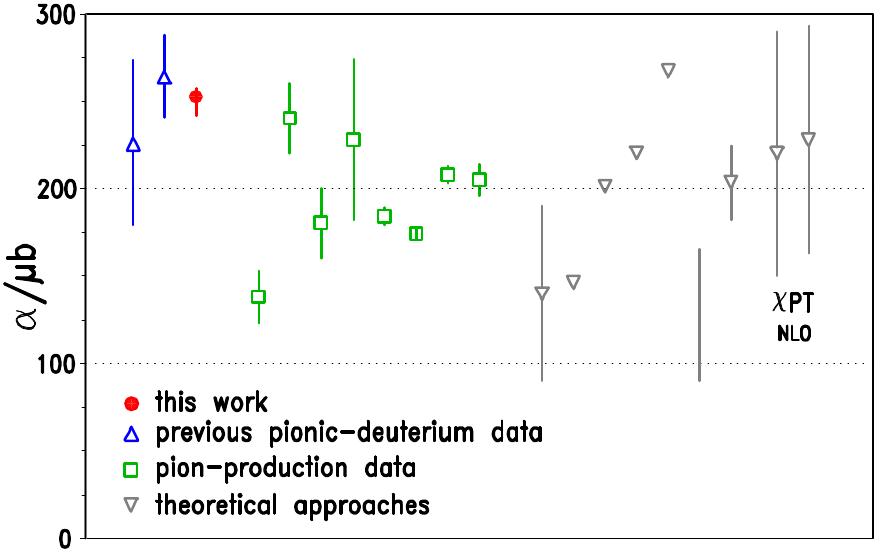}}
\caption{Threshold parameter $\alpha$ for s-wave pion production compared to results from previous experiments and theoretical studies(from left). 
         Previous X-ray measurements: \cite{Hau98}, \cite{Cha9597};
         pion production and absorption: \cite{Cra55}, \cite{Ros67}, \cite{Ric70}, \cite{Aeb76}, \cite{Hut90}, \cite{Rit91} ,\cite{Hei96}, \cite{Dro98};
         Theoretical calculations: \cite{Ros54}, \cite{Kol69}, \cite{Rei69}, \cite{Afn74}, \cite{Afn74}, \cite{Fay80}, \cite{Hor93,Nis96}, 
                                   \cite{Len06}, \cite{Len07}.
         Points are given in the same order as listed in table\,\ref{table:scat}.}
\label{figure:alpha}
\end{center}
\end{figure}

The theoretical understanding of $NN\leftrightarrow NN\pi$ reaction is continuously increasing. Within the 
approach of $\chi$PT, a study of pion production including the terms in NLO yields 
$\alpha^{\mathrm{NLO}}=(220\,\pm\,70)\,\mu$b \cite{Len06} in good agreement with the pionic deuterium results 
(Table\,\ref{table:scat} and fig.\ref{figure:alpha}). The theoretical uncertainty is expected to decrease to 
below 10\% within a few years from forthcoming NNLO calculations\,\cite{Han09}. 

The parameter $\alpha$ when determined from pion production experiments shows wide fluctuations even when 
comparing recent data. Often, only the statistical error is given for the cross section of the production 
experiments, but the fluctuations suggest significant systematic uncertainties, which may arise from uncertainties 
in the normalisation and/or Coulomb corrections. 
The expected order of magnitude of isospin breaking effects is about or less the precision of this experiment, 
which is far below the variation of the pion production data.

\subsection{Pion absorption on nucleon-nucleon pairs}\label{subsec:absorption} 

To describe the s-wave absorption strength  on isoscalar (I=0) and isovector (I=1) nucleon-nucleon pairs ($NN$), 
effective couplings $g_{0}$ and $g_{1}$ have been introduced representing the transitions 
$^{3}S_{1}(I=0)\rightarrow\,^{3}P_{1}(I=1)$ ($g_0$) and $^{1}S_{0}(I=1)\rightarrow\,^{3}P_{0}(I=1)$ 
($g_1$)\,\cite{Eck63}. The transition strength $\mid\,g_{0}\mid^2$ is hence related to the imaginary part of the 
scattering length $a_{\pi\mathrm{D}}$ as determined from $\Gamma_{1s}$ in 
$\pi$D by $\mathrm{Im}\,a_{\pi\mathrm{D}}=\frac{1}{3}\,(p^{*}_{n})^{3}\mid g_0\mid^2(1+1/S')$\,\cite{Ger88} and, 
consequently, also to the threshold parameter $\alpha$ by 
\begin{eqnarray}
\mid\,g_{0}\mid^2=\frac{1}{2\pi m_{\pi}}\cdot\frac{p^{*}_{p}}{p^{*}_{n}}\cdot \frac{1}{p^{*}_{n}}\cdot \alpha 
\end{eqnarray}
(see sect.\,\ref{sec:strong_interaction_width}). The CMS momentum of the neutrons acquired in the absorption 
reaction is $p^{*}_{n}=2.6076\,m_{\pi}$. 

The relative strength of the coupling for absorption on isoscalar to isovector $NN$ pairs was obtained 
to $\mid g_{0}/g_{1}\mid =2.48\,\pm\,0.24$ from the ratio of neutron-neutron to neutron-protons pairs 
emitted back-to-back after pion absorption in the helium isotopes\,\cite{Got95,Dau95}. From $\alpha$ as 
given by $\mathrm{Im}\,a_{\pi\mathrm{D}}$ and $\mid g_{0}/g_{1}\mid$ one obtains 
\begin{eqnarray}
\mid g_{0}\mid &=& \left(2.780{+\,0.027\atop-\,0.058}\right)\cdot 10^{-2}~m^{-2}_{\pi}\\
\mid g_{1}\mid &=& ~\left(1.12\,~\pm~0.11\,\right)\cdot 10^{-2}~m^{-2}_{\pi}\,,
\end{eqnarray}
where the effective coupling $g_{1}$ describes, {\it e.g.}, the threshold transition strength of the reaction 
$pp\rightarrow pp\pi^{0}$\,\cite{Rog89}. 

A phenomenological analysis for $NN\rightarrow NN\pi$ close to threshold using the ansatz 
$\sigma =B_{0}\eta^{2}$\,\cite{Ros54} 
yields from fits to the cross section $B_{0}=27.6\,\mu$b. The uncertainty is estimated to 
be as large as 50\% \cite{Lon82}. However, the simple energy dependence is only valid in the limit of an 
infinitely large scattering length. 

Scaling our result for $g_1$ one obtains a smaller value of 
$B_{0}=13.3\pm 1.4\,\mu$b. This may be due to the fact that the extraction on both $\mid g_{0}/g_{1}\mid$ 
from $\pi^{-}He$ absorption and $g_{0}$ from $\pi$D are affected differently by $NN$ final state interactions. 

As discussed, {\it e.g.} by\,\cite{Mil91,Mey92}, the phenomenological ansatz neglects the strong $NN$ final-state 
interaction leading to a completely different and complicated energy dependence close to threshold, in particular, 
when comparing $pn$ and $pp$ ($^1S_0$) final states\,\cite{Han04,Bar09}. This prevents a direct determination 
of the production amplitude from a simple parametrization of the cross section like $\propto\eta^{2}$. 

Calculations including final-state and Coulomb interaction lead to values for $\sigma/\eta^{2}$ as small as 
5\,$\mu$b when approaching threshold\,\cite{Mil91,Mey92}. This is in agreement with the long known suppression 
of the $pp\rightarrow pp\pi^{0}$ amplitude in LO\,\cite{Kol69} and a $\chi$PT calculation 
showing that loop contributions vanish up to NLO\,\cite{Han02}.

\section{Summary}\label{sec:summary}

The $\pi$D$(3p-1s)$ X-ray transition in pionic deuterium has been studied using a high-resolution crystal 
spectrometer to determine the ground-state strong-interaction effects. 

The accuracy of 1.3\%, obtained for the shift, supersedes the theoretical uncertainty of 9\% achieved 
recently, and which can be traced back mainly to insufficient know\-ledge of low-energy constants. However, when 
used as a constraint for the pion-nucleon scattering lengths $a^+$ and $a^-$ as determined from pionic hydrogen, 
the new $\pi$D result reduces the uncertainty for the isoscalar scattering length $a^+$ by a factor of about 
two yielding a non-zero and positive value\,\cite{Bar11}.

The accuracy achieved for the hadronic broadening of 4.2\% reaches already the expected final uncertainty of 
$(5-10)$\% of forthcoming NNLO $\chi$PT calculations\,\cite{Han09}. From its value, the transition strength 
$\alpha$ for s-wave pion production in nucleon-nucleon collisions is determined with unprecedented accuracy. 

No line broadening by radiative de-excitation from non stabilised molecular states was identified which is 
corroborated from the study of the energy dependence of the $\pi$D$(3p-1s)$ transition energy. However, 
weak evidence (2$\sigma$) was found for a Auger stabilised state.

Noteworthy, that at the 10\% level no high-energy components could be identified stemming from low-lying \linebreak
Coulomb de-excitation transitions in contrast to pionic and muonic hydrogen.

\section*{Acknowledgements}
We would like to thank N.\,Dolfus, B.\,Leoni, L.\,Stohwasser, and K.-P.\,Wieder for the technical assistance 
and V.\,Baru, C.\,Hanhart, A.\,Nogga, M.\,Hoferichter, and A.\,Rusetsky for continuous 
exchange of information on theoretical progress on the $\pi$D system. 
The Bragg crystal was manufactured by Carl Zeiss AG, Oberkochen, Germany. 
Partial funding and travel support was granted by FCT (Lisbon) and FEDER (Grant No. SFRH/BD/18979/2004 and project 
PTDC/FIS/102110/2008) and the Germaine de Sta$\ddot{e}$l exchange program. 
LKB is Unit\'{e} Mixte de Recherche du CNRS, de l'\'{E}cole Normale Sup\'{e}rieure et de UPMC No. 8552. 
This work is part of the PhD thesis of one of us (Th.\,S., Univ. of Cologne, 2009). 
%


\begin{thebibliography}{999}

\bibitem{Des54}  S.\,Deser, M.\,L.\,Goldberger, K.\,Baumann, and W.\,Thirring, Phys. Rev. {\bf 96}, 774 (1954).
\bibitem{Got04}  D.\,Gotta, Prog. Part. Nucl. Phys. {\bf 52}, 133 (2004).
\bibitem{Eri88-6}  T.\,E.\,O.\,Ericson and W.\,Weise, {\em Pions and Nuclei}, (Clarendon, Oxford, 1988) chapter 6.
\bibitem{Bru55}  K.\,A.\,Br\"uckner, Phys. Rev. {\bf 98}, 769 (1955).
\bibitem{Gas08}  J.\,Gasser, V.\,E.\,Lyubovitskij, and A.\,Rusetsky, Phys. Rep. {\bf 456}, 167 (2008).
\bibitem{Lue86}  M.\,L\"uscher, Commun. Math. Phys {\bf 105}, 153 (1986); Nucl. Phys. B {\bf 354}, 531 (1991).
\bibitem{Tor10}  A.\,Torok {\it et al.}, Phys. Rev. D {\bf 81}, 074506 (2010).
\bibitem{Bea11}  S.\,R.\,Beane, W.\,Detmold, K.\,Originos, and M.\,J.\,Savage, Prog. Part. Nucl. Phys. {\bf 66}, 1 (2011).

\bibitem{Wei66}  S.\,Weinberg, Phys. Rev. Lett. {\bf 17}, 616 (1966).
\bibitem{Tom66}  Y.\,Tomozawa,  Nuovo Cim. A {\bf 46}, 707 (1966).
\bibitem{Gas82}  J.\,Gasser and H.\,Leutwyler, Phys. Rep. {\bf 87}, 77 (1982).
\bibitem{ThoWei} A.\,W.\,Thomas and W.\,Weise, {\em The Structure of the Nucleon}, (WILEY--VCH, Berlin, 2001) chapter 7.

\bibitem{Wei79}  S.\,Weinberg, Physica {\bf 96A}, 327 (1979).
\bibitem{Gas84}  J.\,Gasser and H.\,Leutwyler, Ann. Phys. {\bf 158}, 142 (1984).
\bibitem{Gas85}  J.\,Gasser and H.\,Leutwyler, Nucl. Phys. B {\bf 250}, 465 (1985).
\bibitem{Eck95}  G.\,Ecker, Prog. Part. Nucl. Phys. {\bf 35}, 1 (1995).
\bibitem{Ber95}  V.\,Bernard, N.\,Kaiser, and Ulf-G.\,Mei{\ss}ner, Int. J. Mod. Phys. E {\bf 4}, 193 (1995).
\bibitem{Sch02}  S.\,Scherer, {\em Introduction to Chiral Perturbation Theory}, Adv. Nucl. Phys. {\bf 27}, 277 (2003) (arXiv:hep-ph/0210398v1).
\bibitem{Ber08}  V.\,Bernard, Prog. Part. Nucl. Phys. {\bf 60}, 82 (2008).

\bibitem{Bec01}  T.\,Becher and H.\,Leutwyler, J. High En. Phys. {\bf 06}, 017 (2001).
\bibitem{Gas91}  J.\,Gasser, H.\,Leutwyler, and M.\,E.\,Sainio, Phys. Lett. B {\bf 253}, 252 (1991).
\bibitem{Sai02}  M.\,E.\,Sainio, {\em Pion-nucleon $\sigma$-term -- a review},
                 in {\em Proc. of the 9$^{\,th}$ Symp. on Meson-Nucleon Physics and the Structure of the Nucleon\,(MENU'01)},
                 Washington D.\,C., 2001, $\pi N$\,newsletter {\bf 16}, 138 (2002), ISSN 0942-4148, and references therein.
\bibitem{Gol55}  M.\,L.\,Goldberger, H.\,Miyazawa and R.\,Oehme, Phys. Rev. {\bf 99}, 986 (1955).
\bibitem{Eri02}  T.\,E.\,O.\,Ericson, B.\,Loiseau and A.\,W.\,Thomas, Phys. Rev. C {\bf 66}, 014005 (2002).
\bibitem{Aba07}  V.\,V.\,Abaev, P.\,Mets$\mathrm{\ddot{a}}$, and M.\,E.\,Sainio, Eur. Phys. J. A {\bf 32}, 321 (2007).
\bibitem{Ber96}  V.\,Bernard, N.\,Kaiser, and Ulf-G.\,Mei{\ss}ner, Phys. Lett. B {\bf 383}, 116 (1996).
\bibitem{Kov97}  M.\,A.\,Kovash {\it et al.}, $\pi N$ {\it newsletter} {\bf 12}, 51 (1997).
\bibitem{Han97}  O.\,Hanstein, D.\,Drechsel, and L.\,Tiator, $\pi N$ {\it newsletter} {\bf 12}, 56 (1997).
\bibitem{Mea01}  D.\,F.\,Measday, Phys. Rep. {\bf 354}, 243 (2001) .
\bibitem{Gor04}  T.\,Gorringe and H.\,W.\,Fearing, Rev. Mod. Phys. {\bf 76}, 31 (2004).
\bibitem{And07}  V.\,A.\,Andreev {\it et al.}, Phys. Rev. Lett. {\bf 99}, 032002 (2007).

\bibitem{Sig96b} D.\,Sigg {\it et al.}, Nucl. Phys. A {\bf 609}, 310 (1996).
\bibitem{Lyu00}  V.\,E.\,Lyubovitskij and A.\,Rusetsky, Phys. Lett. B {\bf 494}, 9 (2000).
\bibitem{Gas02}  J.\,Gasser {\it et al.}, Eur. Phys. J. C {\bf 26}, 13 (2002).
\bibitem{Zem03}  P.\,Zemp, {\em Proc. of Chiral Dynamics 2003}, p.\,128, Bonn, Germany, September 8--13, 2003, arXiv:hep-ph/0311212v1; 
                 {\it Pionic Hydrogen in QCD+QED: Decay width at NNLO}, Ph.\,D. Thesis, University of Bern (2004).
\bibitem{Ras82}  G.\,Rasche and W.\,S.\,Woolcock, Nucl. Phys., A {\bf 381}, 405 (1982).
\bibitem{Spu77}  J.\,Spuller {\it et al.}, Phys. Lett. {\bf 67} B, 479 (1977).
\bibitem{Oad07}  G.\,C.\,Oades, G.\,Rasche, W.S.\,Woolcock, E.\,Matsinos, A.\,Gashi, Nucl. Phys. A {\bf 794}, 73 (2007).
\bibitem{Eri88-9}  T.\,E.\,O.\,Ericson and W.\,Weise, {\em Pions and Nuclei}, (Clarendon, Oxford, 1988) chapter 9.
\bibitem{Eri88-4} {\it ibidem}, chapter 4.

\bibitem{ThLa80} A.\,W.\,Thomas and R.\,H.\,Landau, Phys. Rep. B {\bf 58}, 121 (1980).
\bibitem{Del03}  A.\,Deloff, {\em Fundamentals in Hadronic Atom Theory}, (World Scientific, London, 2003) chapter 15.

\bibitem{Sig96a} D.\,Sigg {\it et al.}, Nucl. Phys. A {\bf 609}, 269 (1996).
\bibitem{Cha9597}  D.\,Chatellard {\it et al.}, Phys. Rev. Lett. {\bf 74}, 4157 (1995); Nucl. Phys. A {\bf 625}, 855 (1997).
\bibitem{Hau98}  P.\,Hauser {\it et al.}, Phys. Rev. C {\bf 58}, R1869 (1998).
\bibitem{Sch01}  H.-Ch.\,Schr\"oder {\it et al.}, Eur. Phys. J C {\bf 21}, 473 (2001).

\bibitem{Wei92}  S.\,Weinberg, Phys. Lett. B {\bf 295}, 114 (1992).
\bibitem{Bar97}  V.\,Baru and A.\,E.\,Kudryavtsev, Phys. of At. Nucl. {\bf 60}, 1475 (1997).
\bibitem{Bea98}  S.\,R.\,Beane, V.\,Bernard, T.-S.\,Lee, and U.-G.\,Mei{\ss}ner, Phys. Rev. C {\bf 57}, 424 (1998).
\bibitem{Tar00}  V.\,E.\,Tarasov, V.\,V.\,Baru, and A.\,E.\,Kudryavtsev, Phys. At. Nucl. {\bf 63}, 801 (2000).
\bibitem{Del01}  A.\,Deloff, Phys. Rev. C {\bf 64} (2001) 065205.
\bibitem{Kai02}  N.\,Kaiser, Phys. Rev. C {\bf 65}, 057001 (2002).
\bibitem{Bur03}  B.\,Burasoy and H.\,W.\,Grieshammer, Int. J. Mod. Phys. E {\bf 12}, 65 (2003).
\bibitem{Bea03}  S.\,R.\,Beane, V.\,Bernard, E.\,Epelbaum, U.-G.\,Mei{\ss}ner, and D.\,R.\,Phillips,
                 Nucl. Phys. A {\bf 720}, 399 (2003).
\bibitem{Doe04}  M.\,D\"{o}ring, E.\,Oset, and M.\,J.\,Vicente~Vacas, Phys. Rev. C {\bf 70}, 045203 (2004).
\bibitem{Irg04}  B.\,F.\,Irgaziev and B.\,A.\,Fayzullaev, arXiv:hep-ph/0404203v1 (2004).
\bibitem{Mei05}  U.-\,G.\,Mei{\ss}ner, U.\,Raha, and, A.\,Rusetsky, Eur. Phys. J. C {\bf 41}, 213 (2005);
                 Eur. Phys. J. C {\bf 45}, 545 (2006).
\bibitem{Mei06}  U.-\,G.\,Mei{\ss}ner, U.\,Raha, and, A.\,Rusetsky, Phys. Lett. B {\bf 639}, 478 (2006).
\bibitem{Hof09a} M.\,Hoferichter, B.\,Kubis and U.-G.\,Mei{\ss}ner, Phys. Lett. B {\bf 678}, 65 (2009); 
                 Nucl. Phys. A {\bf 833},18 (2010).
\bibitem{Bar11}  V.\,Baru {\it et al.}, Phys. Lett. B {\bf 694}, 473 (2011) (arXiv:nucl-th/1003.4444v2).
\bibitem{Hof09b} M.\,Hoferichter, B.\,Kubis and U.-G.\,Mei{\ss}ner, PoS {\bf CD09}, 014 (2009) (arXiv:hep-ph/0910.0736).

\bibitem{Hue75}  J.\,H\"ufner, Phys. Rep. {\bf 21}, 1 (1975).
\bibitem{Len06}  V.\,Lensky {\it et al.}, Eur. Phys. J. A {\bf 27}, 37 (2006).
\bibitem{Len07}  V.\,Lensky {\it et al.}, Phys. Lett. B {\bf 648}, 46 (2007).

\bibitem{Hig81}  V.\,L.\,Highland {\it et al.}, Nucl. Phys. A {\bf 365}, 333 (1981).
\bibitem{Jos60}  D.\,W.\,Joseph, Phys. Rev. {\bf 119}, 805 (1960).
\bibitem{Don77}  R.\,MacDonald {\it et al.}, Phys. Rev. Lett. {\bf 38}, 746 (1977).

\bibitem{PSI98}  PSI proposal R-98-01, www2.fz-juelich.de/ikp/exotic-atoms.
\bibitem{Got08}  D.\,Gotta {\it et al.}, Lect. Notes Phys. {\bf 745}, 165 (2008).
\bibitem{Str10}  Th.\,Strauch {\it et al.}, Phys. Rev. Lett. {\bf 104}, 142503 (2010).

\bibitem{Har90}  F.J.\,Hartmann, {\it Proceedings of Physics of Exotic Atoms on Electromagnetic Cascade and Chemistry}, 
                 Erice, Italy, 1989 (Plenum Press, New York, 1990) p.\,23 and p.\,127, and references therein.
\bibitem{Coh04}  J.S.\,Cohen, Rep. Prog. Phys. {\bf 67}, 1769 (2004).
\bibitem{Bor80}  E.\,Borie and M.\,Leon, Phys. Rev. A {\bf 21}, 1460 (1980).
\bibitem{Leo62}  M.\,Leon and H.~A.~Bethe, Phys. Rev. {\bf 127}, 636 (1962).
\bibitem{Has95}  R.\,al\,Hassani {\it et al.}, Z. Phys. A {\bf 351}, 113 (1995).
\bibitem{Jen02a} T.\,S.\,Jensen and V.\,E.\,Markushin, Eur. Phys. J. D {\bf 19}, 165 (2002).
\bibitem{Jen02b} T.\,S.\,Jensen and V.\,E.\,Markushin, Eur. Phys. J. D {\bf 21}, 261 (2002).
\bibitem{Jen02c} T.\,S.\,Jensen and V.\,E.\,Markushin, Eur. Phys. J. D {\bf 21}, 271 (2002).
\bibitem{BF78}   L.\,Bracci and G.\,Fiorentini, Nuovo Cim. A {\bf 43}, 9 (1978).
\bibitem{Poh06}  R.\,Pohl {\it et al.}, Phys. Rev. Lett. {\bf 97}, 193402 (2006).
\bibitem{Czi63}  J.\,B.\,Czirr {\it et al.}, Phys. Rev. {\bf 130}, 341 (1963).
\bibitem{Bad01}  A.\,Badertscher {\it et al.}, Europhys. Lett. {\bf 54}, 313 (2001), and references therein.
\bibitem{Cov09}  D.\,S.\,Covita {\it et al.}, Phys. Rev. Lett. {\bf 102}, 023401 (2009).
\bibitem{Hen03}  M.\,Hennebach, {\it Precision measurement of ground state transitions in pionic hydrogen}, 
                 Ph.\,D. thesis, Universit\"at zu K\"oln (2003), http://kups.ub.uni-koeln.de/744/.
\bibitem{PP06}   V.\,N.\,Pomerantsev and V.\,P.\,Popov, Phys. Rev. A {\bf 73}, 040501(R) (2006).
\bibitem{JPP07}  T.\,S.\,Jensen, V.\,N.\,Pomerantsev, and V.\,P.\,Popov, arXiv:nucl-th/0712.3010v1 (2007).
\bibitem{PP07}   V.\,P.\,Popov and V.\,N.\,Pomerantsev, arXiv:nucl-th/0712.3111v1 (2007).
\bibitem{Taq89}  D.\,Taqqu, AIP Conf. Proc. {\bf 181}, 217 (1989).
\bibitem{Jon99}  S.\,Jonsell, J.\,Wallenius, P.\,Froelich, Phys. Rev. A {\bf 59}, 3440 (1999).
\bibitem{Poh09}  R.\,Pohl, Hyperfine Int. {\bf 193}, 115 (2009).
\bibitem{Lin03}  E.\,Lindroth, J.\,Wallenius, S.\,Jonsell, Phys. Rev. A {\bf 68}, 032502 (2003); Phys. Rev. A {\bf 69}, 059903(E) (2004).
\bibitem{Kil04}  S.\,Kilic, J.-P.\,Karr, L.\,Hilico, Phys. Rev. A {\bf 70}, 042506 (2004).

\bibitem{Tru61}  T.\,L.\,Trueman, Nucl. Phys. {\bf 26}, 57 (1961).
\bibitem{Lam6970}  E.\,Lambert, Helv. Phys. Acta {\bf 42} (1969) 667; {\em ibidem} {\bf 43}, 713 (1970).
\bibitem{Mit01}  J.\,Mitroy and I.\,A.\,Ivallov, J. Phys. G {\bf 27}, 1421 (2001).
\bibitem{Eir00}  D.\,Eiras and J.\,Soto, Phys. Lett. B {\bf491}, 101 (2000).

\bibitem{Ros54}  A.\,H.\,Rosenfeld, Phys. Rev. {\bf 96}, 139 (1954).
\bibitem{Rei69}  A.\,Reitan, Nucl. Phys. B {\bf 11}, 170 (1969).
\bibitem{Mac06}  H.\,Machner and J.\,Niskanen, Nucl.Phys. A {\bf 776}, 172 (2006).
\bibitem{Bru51}  K.\,Br\"uckner, R.\,Serber, and K.\,Watson, Phys. Rev. {\bf 81}, 575 (1951).

\bibitem{Barpc}  V.\,Baru, C.\,Hanhart, and A.\,Rusetski, priv. comm.
\bibitem{Fil09}  A.\,Filin {\it et al.}, Phys. Lett. B {\bf 681}, 423 (2009).
\bibitem{Byc73}  E.\,Byckling and K.\,Kajantie, {\em Particle Kinematics}, John Wiley and Sons, London (1973) chapter II.6.


\bibitem{Sim88} L.\,M.\,Simons, Physica Scripta {\bf T22}, 90 [1988).
\bibitem{Sim93} L.\,M.\,Simons, Hyperfine Int. {\bf 81}, 253 (1993).
\bibitem{Cov08}  D.\,S.\,Covita {\it et al.}, Rev. Scient. Instr. {\bf 79}, 033102 (2008).
\bibitem{Egg65}  J.\,Eggs and K.\,Ulmer, Z. angew. Phys., {\bf 20(2)}, 118 (1965).
\bibitem{Zsc82}  G.\,Zschornack, Nucl. Instr. Meth. {\bf 200}, 481 (1982).
\bibitem{Nel02}  N.\,Nelms {\it et al.}, Nucl. Instr. Meth. A {\bf 484}, 419 (2002).

\bibitem{Bas94}  G.\,Basile {\it et al.}, Phys. Rev. Lett. {\bf 72}, 3133 (1994).
\bibitem{Des03}  R.\,Deslattes {\it et al.}, Rev. Mod. Phys., vol. {\bf  75}, no. 1, 35 (2003).
\bibitem{Ind06}  P.\,Indelicato {\it et al.}, Rev. Sci. Instrum. {\bf 77}, 043107 (2006).
\bibitem{Ana05}  D.\,F.\,Anagnostopulos {\it et al.}, Nucl. Instr. Meth. A {\bf 545}, 217 (2005).
\bibitem{Tra07}  M.\,Trassinelli {\it et al.}, J. Phys., Conf. Ser. {\bf 58}, 129 (2007).
\bibitem{Covth}  D.\,S.\,Covita, {\it High-precision spectroscopy of the 3p -- 1s transition in muonic hydrogen}, 
                 Ph.\,D. thesis, University of Coimbra (2008), http://hdl.handle.net/10316/7521.
\bibitem{San98}  M.\,Sanchez del Rio and R.\,J.\,Dejus, Proc. SPIE Int. Soc. Opt. Eng. {\bf 3448}, 246 (1998); 
                 {\it ibidem} {\bf 5536}, 171 (2004); http://www.esrf.eu/computing/scientific/xop2.1.
\bibitem{Str09}  Th.\,Strauch, {\it High-precision measurement of strong-interaction effects in pionic deuterium}, 
                 Ph.\,D. thesis, Universit\"at zu K\"oln (2009), http://kups.ub.uni-koeln.de/2813/.

\bibitem{Vei73}  WM.\,J.\,Veigele, At. Data Tables {\bf 5}, 51 (1973).
\bibitem{Kra79}  M.\,O.\,Krause and J.\,H.\,Oliver, J. Phys. Chem. Ref. Data {\bf 8}, 329 (1979).
\bibitem{Moo07}  T.\,Mooney, Argonne National Laboratory, priv. comm. (2007).
\bibitem{Hen93}  B.\,L.\,Henke, E.\,M.\,Gullikson, and J.\,C.\,Davies, At. Data Nucl. Data Tables {\bf 54}, 181 (1993).
\bibitem{Cha95}  C.\,T.\,Chantler, J. Phys. Chem. Ref. Data {\bf 24}, 71 (1995).
\bibitem{Chu96}  F.\,N.\,Chukhovskii, G.\,H\"olzer, O.\,Wehrhan, and E.\,F\"orster, J. Appl. Cryst. {\bf 29}, 438 (1998).
\bibitem{Cem92}  F.\,Cembali {\it et al.}, J. Appl. Cryst. {\bf 24}, 424 (1992).
\bibitem{Jam75}  F.\,James and M.\,Roos, Comput. Phys. Commun. {\bf 10}, 343 (1975).
\bibitem{Ber02}  U.\,C.\,Bergmann and K.\,Riisager, Nucl. Instrum. Meth. A {\bf 489}, 444 (2002).

\bibitem{PDG08}  C.\,Amsler {\it et al.} (PDG 2008), Phys. Lett. B {\bf 667}, 1 (2008).
\bibitem{Poh10}  R.\,Pohl {\it et al.}, Nature {\bf 466}, 213 (2010).
\bibitem{Par10}  C.\,G.\,Parthey {\it et al.}, Phys. Rev. Lett. {\bf 104}, 233001 (2010).
\bibitem{Ros93}  Y.\,Lu and R.\,Rosenfelder, Phys. Lett. B {\bf 319}, 7 (1993); Phys. Lett. B {\bf 333}, 564(E) (1994).
\bibitem{Ros95}  W.\,Leidemann and R.\,Rosenfelder, Phys. Rev. B {\bf 51}, 427 (1995).
\bibitem{Ahr05}  J.\,Ahrens {\it et al.}, Eur. Phys. J. A {\bf 23}, 113 (2005).
\bibitem{Koc86}  R.\,Koch, Nucl. Phys. A {\bf 448}, 707 (1986).

\bibitem{Bai74}  J.\,Bailey {\it et al.}, Phys. Lett. {\bf 50} B, 403 (1974).
\bibitem{Bov81}  E.\,Bovet {\it et al.}, Nucl. Instr. Meth. {\bf 190}, 613 (1981).
\bibitem{Bov85}  E.\,Bovet {\it et al.}, Phys. Lett. {\bf 153} B, 231 (1985).

\bibitem{Han09}  C.\,Hanhart, priv. comm. (2010).

\bibitem{Cra55}  F.\,S.\,Crawford and M.~L.~Stevenson, Phys. Rev. {\bf 97}, 1305 (1955).
\bibitem{Ros67}  C.\,M.\,Rose, Phys. Rev. {\bf 154}, 1305 (1967).
\bibitem{Ric70}  C.\,Richard-Serre {\it et al.}, Nucl. Phys. B {\bf 20}, 413 (1970).
\bibitem{Aeb76}  D.\,Aebischer {\it et al.}, Nucl. Phys. B {\bf 108}, 214 (1976).
\bibitem{Hut90}  D.\,A.\,Hutcheon {\it et al.}, Phys. Rev. Lett. {\bf 64}, 176 (1990); Nucl. Rev. A {\bf 535}, 618 (1991).
\bibitem{Rit91}  B.\,G.\,Ritchie {\it et al.}, Phys. Rev. Lett. {\bf 66}, 568 (1991).
\bibitem{Hei96}  P.\,Heimberg {\it et al.}, Phys. Rev. Lett. {\bf 77}, 1012 (1996).
\bibitem{Dro98}  M.\,Drochner {\it et al.}, Nucl. Phys. A {\bf 643}, 55 (1998).
\bibitem{Kol69}  D.\,S.\,Koltun and A.\,Reitan, Phys. Rev. {\bf 141}, 1413 (1966).
\bibitem{Afn74}  I.\,R.\,Afnan and A.\,W.\,Thomas, Phys. Rev. C {\bf 10}, 109 (1974).
\bibitem{Fay80}  C.\,Fayard, G.\,H.\,Lamot, and T.\,Mizutani, Phys. Rev. Lett. {\bf 45}, 524 (1980).
\bibitem{Hor93}  C.\,J.\,Horowitz, Phys. Rev. C {\bf 48}, 2920 (1993).
\bibitem{Nis96}  J.\,A.\,Niskanen, Phys. Rev. C {\bf 53}, 526 (1996).
\bibitem{Iva97}  A.\,N.\,Ivanov {\it et al.}, Z. Phys. A {\bf 358}, 81 (1997).

\bibitem{Eck63}  S.\,G.\,Eckstein, Phys. Rev. {\bf 129}, 413 (1963).
\bibitem{Ger88}  J.-F. Germond and C.\,Wilkin, J. Phys. G {\bf 14}, 181 (1988).
\bibitem{Got95}  D.\,Gotta {\it et al.}, Phys. Rev. C {\bf 51}, 469 (1995).
\bibitem{Dau95}  E.\,Daum {\it et al.}, Nucl. Phys. A {\bf 589}, 553 (1995).
\bibitem{Rog89}  J.\,Roginsky and C.\,Werntz, Phys. Rev. C {\bf 40}, 2732 (1989).
\bibitem{Lon82}  D.\,G.\,Long, M.\,Sternheim, and R.\,R.\,Silbar, Phys. Rev  C {\bf 26}, 586 (1982).
\bibitem{Mil91}  G.\,A.\,Miller and P.\,Sauer, Phys. Rev. C {\bf 44}, R1725 (1991).
\bibitem{Mey92}  H.\,O.\,Meyer {\it et al.}, Nucl. Phys. A {\bf 539}, 633 (1992).
\bibitem{Han04}  C.\,Hanhart, Phys. Rep. {\bf 397}, 155 (2004).
\bibitem{Bar09}  V.\,Baru {\it et al.},  Phys. Rev. C {\bf 80}, 044003 (2009).
\bibitem{Han02}  C.\,Hanhart and N.\,Kaiser, Phys. Rev. C {\bf 66}, 054005 (2002).



\end{thebibliography}
\end{document}